\documentclass[a4paper,12pt,twoside]{article}
\usepackage{psfig}
\usepackage{epsfig}
\usepackage{amssymb,wasysym,bbm,feynmf,rotating,subfigure,here,citesort}
\usepackage{graphicx}
\newcommand{\be}{\begin{equation}}
\newcommand{\ee}{\end{equation}}
\newcommand{\bea}{\begin{eqnarray}}
\newcommand{\eea}{\end{eqnarray}}
\newcommand{\benn}{\begin{displaymath}}
\newcommand{\eenn}{\end{displaymath}}
\newcommand{\beann}{\begin{eqnarray*}}
\newcommand{\eeann}{\end{eqnarray*}}
\newcommand{\gtsim}{\lower-0.45ex\hbox{$>$}\kern-0.77em\lower0.55ex\hbox{$\sim$}}
\newcommand{\ltsim}{\lower-0.45ex\hbox{$<$}\kern-0.77em\lower0.55ex\hbox{$\sim$}}
\newcommand{\befig}{\begin{figure}}
\newcommand{\efig}{\end{figure}}
\newcommand{\betab}{\begin{table}}
\newcommand{\etab}{\end{table}}
\begin{document}
%
%
% ___ The Title Page ________________________________________________
%
\pagestyle{empty}
\title{ \vspace*{-1cm} {\normalsize\rightline{CU-TP-1104}}
  \vspace*{0.cm} 
{\Large \bf
    \boldmath Small-$x$ physics beyond the Kovchegov
  equation}\footnote{This work is supported in part by the US
  Department of Energy.}}

\author{}
\date{} \maketitle
 
\vspace*{-2.5cm}
 
\begin{center}
 
\renewcommand{\thefootnote}{\alph{footnote}}
 
{\large
A.~H.~Mueller\,\footnote{arb@phys.columbia.edu (A.H.Mueller)} and
A.~I.~Shoshi\,\footnote{shoshi@phys.columbia.edu}}
 
%\vspace*{0.5cm}
 
{\it Physics Department, Columbia University, New York, NY 10027, USA}

%\medskip
 
\end{center}
 
% Text of abstract

\begin{abstract}
  
  We note the differences between the Kovchegov equation and the
  Balitsky-JIMWLK equations as methods of evaluating high energy hard
  scattering near the unitarity limit. We attempt to simulate some of
  the correlations absent in the Kovchegov equation by introducing two
  boundaries rather than the single boundary which effectively
  approximates the unitarity limit guaranteed in the Kovchegov
  equation. We solve the problem of BFKL evolution in the presence of
  two boundaries and note that the resulting $T$-matrix now is the
  same in different frames, which was not the case in the single
  boundary case. The scaling behavior of the solution to the
  Kovchegov equation is apparently now lost.

%\bigskip
  \vspace{1.cm}
 
% keywords here
 
\medskip

% PACS codes here

\end{abstract}

%%%%%%%%%%%%%%%%%%%%%%%%%%%%%%%%%%%%%%%%%%%%
% ___ Table of Contents _____________________________________________
%
%\newpage
%\pagestyle{empty}
%
%\tableofcontents
%
%\addtocontents{toc}{\protect\enlargethispage{2.cm}}
%%%%%%%%%%%%%%%%%%%%%%%%%%%%%%%%%%%%%%%%%%%%%%%%%%%%

\pagenumbering{roman}
\pagestyle{plain}
%
%
%%%%%%%%%%%%%%%%%%%%%%%%%%%%%%%%%%%%%%%%%%%%%%%%%%%%%%%%%%%%%%%%%%%%
% ___ The Sections __________________________________________________
%
\pagenumbering{arabic}
\pagestyle{plain}
%
% The lines below are necessary in order to enumerate the equations
% according to the sections where they are.
\makeatletter
\@addtoreset{equation}{section}
\makeatother
\renewcommand{\theequation}{\thesection.\arabic{equation}}

%_____ The Introduction ___________________________
\newpage
\section{Introduction}
\label{Sec_Introduction}
% ______________________________________________________________________________
Scattering at the unitarity limit and parton
saturation~\cite{Gribov:tu} are dual descriptions of the same
phenomenon~\cite{Kovchegov:1997dm} which lies at the heart of some of
the most interesting parts of small-$x$ hard scattering. The QCD
dynamics giving the growth of the cross section with increasing
energy, or equivalently, giving the growth of the parton density with
decreasing $x$ is BFKL~\cite{Kuraev:fs,Balitsky:ic} evolution.
Ordinary QCD, or
DGLAP~\cite{Dokshitzer:sg,Gribov:ri,Altarelli:1977zs}, evolution also
leads to increasing parton number densities, however, in order to get
large gluon occupation numbers characteristic of gluon saturation
(Color Glass Condensate~\cite{CGC}) it is necessary that parton
densities, or cross sections, grow at a fixed hard scale and this
requires BFKL evolution. Understanding QCD wavefunctions having high
gluon occupation number and high energy hard scattering near the
unitarity limit are two seemingly different aspects of the same
phenomenon. The partonic language appears to be the more interesting
description because it suggests one should be able to produce quite
dense QCD matter. Indeed, in a central collision of high energy heavy
ions the system produced just after the collision is believed to be a
system of nonequilibrium gluons at high density and high occupation
numbers. This system then evolves into an equilibrated quark-gluon
plasma of lower density and lower occupation
numbers~\cite{Baier:2000sb+x}. 

Of course genuine BFKL evolution is only accurate when gluon
occupation numbers are not too large, or equivalently, when $S$-matrix
elements are not too close to the unitarity limit. The Balitsky
equation~\cite{Balitsky:1995ub+X} is a generalization of the BFKL
equation which should be the leading term in a systematic treatment of
the dynamics of high energy hard scattering at or near the unitarity
limit. An equivalent formalism was developed by Jalilian-Marian, Iancu, McLerran, Leonidov and Kovner
(JIMWLK)~\cite{CGC,Jalilian-Marian:1997jx+X,Iancu:2001ad+X,Weigert:2000gi}
in terms of small-$x$ evolution of QCD wavefunctions in a lightcone
gauge. The Balitsky and JIMWLK equations are coupled equations
involving higher and higher correlations much as in the
Schwinger-Dyson equations in ordinary perturbation theory, and as such
are very difficult to deal with analytically. However, an interesting
numerical calculation has recently appeared~\cite{Rummukainen:2003ns}
and one can expect a better understanding of Balitsky and JIMWLK
evolution to emerge from such calculations. 

Kovchegov~\cite{Kovchegov:1999yj+X} has suggested a somewhat simpler
equation than the Balitsky or JIMWLK equations to deal with scattering
at or near the unitarity limit. Although Kovchegov originally derived
his equation in the context of a high energy scattering on a large
nucleus, this equation can also be viewed as a mean field version of
the Balitsky equation in which higher correlations are neglected. For
example the $S$-matrix for the scattering of a state of two QCD
dipoles on a highly evolved target, and at a definite impact
parameter, is replaced by the product of the $S$-matrices of the
individual dipoles. While the Kovchegov equation is not so complete it
does have the advantage of being a precise nonlinear equation for a
function. Many interesting limits of the Kovchegov equation have been
understood by analytical methods. While incomplete, as is any mean
field like approximation, the Kovchegov equation is likely the best
equation one can write down in terms of functions which has built in
correct unitarity limits for high energy scattering. In
Sec.~\ref{Sec_The Kovchegov_Equation_and_its_Limitations} we give a
derivation of the Kovchegov equation, since its understanding plays a
central role in this paper.

In kinematical regions where the scattering of a QCD dipole on a
target is far from its unitarity limit the Kovchegov equation reduces
to the BFKL equation. At a given impact parameter the dipole target
$T$-matrix ($T=1-S$) grows with energy until it approaches $1$ at
which point the nonlinear part of the Kovchegov equation limits the
growth of $T$ not to be larger than $1$. If $2/Q$ is the transverse
size of the scattering dipole then the saturation momentum is defined
to be the value of $Q$ at which $T$ is some preassigned number, say
$1/2$. This saturation momentum is a function of rapidity $Y$,
$Q_s(Y)$, and gives the scale separating weak and strong scattering.
The transition between weak and strong scattering is expected to be
rapid as one varies the scale $Q$. In the weak scattering region the
BFKL equation can be viewed, roughly, as a product of two factors: The
first gives an exponential dependence in $\rho=\ln(Q^2/\Lambda^2)$
while the second factor represents a diffusion in $\rho$ with $\alpha
Y$ playing the role of time. In Refs.~\cite{Mueller:2002zm} and
\cite{Triantafyllopoulos:2002nz} it was suggested that one could get
an approximate analytic solution to the Kovchegov equation simply by
taking the diffusive part of the BFKL equation and imposing an
absorptive boundary near $\rho_s=\ln(Q^2_s/\Lambda^2)$. This procedure
was then used to evaluate in detail the $Y$-dependence of $Q_s$ and
the form of $T(\rho,Y)$ when $\rho>\rho_s$. These results confirmed
the broad picture arrived at by less complete techniques
earlier~\cite{Gribov:tu,Mueller:1999wm,Iancu:2002tr,Golec-Biernat:2001if}.
Recently Munier and Peschanski~\cite{Munier:2003vc+X} have developed a
more general and more powerful procedure for studying the properties
of the Kovchegov equation near the saturation boundary, confirming the
corrections of the absorptive boundary approach of
Refs.~\cite{Mueller:2002zm,Triantafyllopoulos:2002nz} along the way.

However, the issue still remains to what extend the Kovchegov
equation actually represents the Balitsky and JIMWLK equations at or
near the saturation region. One exact result of the Kovchegov equation
which has been known for some time is the Levin-Tuchin
formula~\cite{Levin:1999mw} for the $S$-matrix deep in the saturation
region which states
\be
S(\rho,b,Y) \sim e^{-c (\rho-\rho_s)^2}
\label{eq_S_c}
\ee
where the constant $c=-C_F(1-\lambda_0)/(N_c 2 \chi(\lambda_0))$ has a
value which follows from the Kovchegov equation. Recently, the authors
of Ref.~\cite{Iancu:2003zr} have claimed that $S$ deep in the
saturation regime has the form given by (\ref{eq_S_c}) but with a
constant at least a factor of $2$ smaller than the $c$ which follows
from the Kovchegov equation. The cause for this discrepancy is the
lack of fluctuations in the Kovchegov equation. Even earlier,
Salam~\cite{Salam:1995uy} has argued on the basis of numerical
simulations that fluctuations are generally large in BFKL evolution
and so there is no reason to expect that they will be small near the
saturation boundary.

\begin{figure}[h!]
\setlength{\unitlength}{1.cm}
\begin{center}
\epsfig{file=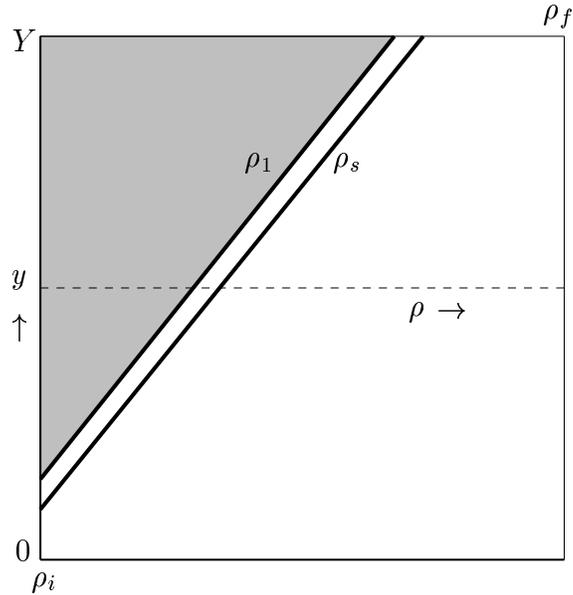, width=7.5cm}
\end{center}
\caption{BFKL evolution in the presence of one saturation boundary.}
\label{Fig_1b}
\end{figure}

It is not hard to see that fluctuations are potentially important in
evolution not too far from the saturation boundary. As an example
consider the high-energy scattering of a dipole of size $\rho_f$ at
rapidity $Y$ on a dipole of size $\rho_i$ at zero rapidity. (The
relation between $\rho$ and the transverse size $x_{\!\perp}$ of a
dipole is $\rho = \ln(4/\Lambda^2 x^2_{\!\perp})$.) We may view the
problem in terms of BFKL evolution with $\rho>\rho_s(y)$ with the
boundary near $\rho_s(y)$ constituting an absorptive boundary for
diffusion inherent in BFKL evolution. Further the evolution can be
viewed as an evolution from ($\rho_i$, $0$) to an intermediate point
($\rho$, $y$) and then from ($\rho$, $y$) to ($\rho_f$, $Y$).  We 
expect the final answer~\cite{Iancu:2002tr} to be close to $T(\rho_i
\to \rho_f) \sim e^{-(1-\lambda_0)(\rho_f-\rho_s(Y))}$. In terms of
the evolution through the intermediate point ($\rho$, $y$) we can
write
\be
T(\rho_i \to \rho_f) \sim \int_{\rho_1(y)}^{\infty}d\rho\,
T(\rho_i \to \rho)\, T(\rho \to \rho_f)
\label{eq_T_to_TT}
\ee
at least as far as the exponential factors are concerned. The $T$'s in
(\ref{eq_T_to_TT}) are all defined by BFKL evolution with an
absorbtive boundary in $\rho$ at $\rho_1(y)$, near $\rho_s(y)$.
Fig.~\ref{Fig_1b} shows in the $\rho$ - $Y$ plane the boundaries
and the shaded saturation region limiting the BFKL evolution.
Clearly $T(\rho_i \to \rho) \sim e^{-(1-\lambda_0)(\rho-\rho_s(y))}$
so one would expect that values of $\rho$ in (\ref{eq_T_to_TT}) such
that $\rho-\rho_s(y)>\rho_f-\rho_s(Y)$ would contribute little to the
integral on the right hand side of (\ref{eq_T_to_TT}). This should be
the case since according to unitarity $T(\rho \to \rho_f)$ should not
be larger than $1$, however, such is not the case. The dominant
contribution to the right-hand side of (\ref{eq_T_to_TT}) comes from
regions with $\rho-\rho_s(y)>\rho_f-\rho_s(Y)$ where manifestly
$T(\rho \to \rho_f) \gg 1$. This occurs because the boundary for
$T(\rho \to \rho_f)$ is matched to the boundary for $T(\rho_i \to
\rho)$, as shown in Fig.~\ref{Fig_1b}, in order to get the same answer
from both sides of eq.~(\ref{eq_T_to_TT}). There is unitarity
violating evolution also in the region with
$\rho-\rho_s(y)<\rho_f-\rho_s(Y)$ when the evolution goes from one
intermediate point to another before it ends at the final point. For
(\ref{eq_T_to_TT}) to be a correct formula and in order to guarantee
unitarity respecting amplitudes for each path of evolution the
saturation boundary for $T(\rho \to \rho_f)$ should follow the value
of $\rho$ and thus it should not be the same as for $T(\rho_i \to
\rho)$. In this case, the BFKL evolution with a saturation boundary
does not fulfill (\ref{eq_T_to_TT}).

\begin{figure}[h!]
\setlength{\unitlength}{1.cm}
\begin{center}
\epsfig{file=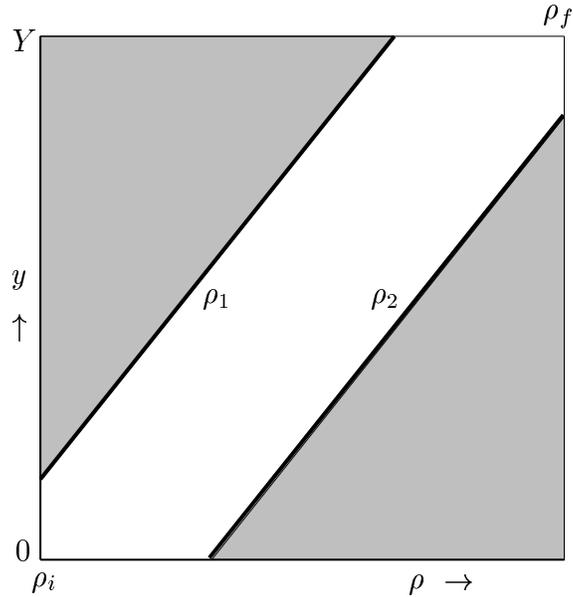, width=7.5cm}
\end{center}
\caption{BFKL evolution in the presence of two saturation boundaries.}
\label{Fig_int_2b}
\end{figure}

In this work we introduce a second absorptive barrier at
$\rho=\rho_2(y)$, as shown in Fig.~\ref{Fig_int_2b}, such that,
crudely, $T(\rho \to \rho_f) \leq 1$ when $\rho<\rho_2$. This second
boundary eliminates all unitarity violating evolution between the
initial point ($\rho_i$, $0$) and the final point ($\rho_f$, $Y$) when
the evolution is viewed as proceeding in two steps; from the initial
point ($\rho_i$, $0$) to an intermediate point ($\rho$, $y$) and then
from the intermediate point ($\rho$, $y$) to the final point
($\rho_f$, $Y$).  However, when the evolution is viewed as proceeding
through two or more intermediate steps, $(\rho_i, 0) \to (\rho, y) \to
(\rho',y'>y) ... \to (\rho_f, Y)$, then the second boundary
$\rho_2(y)$ does not eliminate all unitarity violating evolution
between one intermediate point and another, i.e., $T(\rho \to \rho')$
may become larger than $1$. The remaining unitarity violating
evolution cannot be eliminated without going back to the Balitsky or
JIMWLK equations.

The boundary $\rho_1$ is the boundary corresponding to saturation in
the wavefunction of the evolved dipole $\rho_i$. This boundary is
naturally viewed as an approximate way of imposing the unitarity limit
given by the Kovchegov equation when the scattering is viewed as the
scattering of an elementary dipole $\rho_f$ on a highly evolved dipole
$\rho_i$. However, we may also view the process as the scattering of
an elementary dipole $\rho_i$ on the evolved dipole $\rho_f$. When
$\rho_f>\rho_i$ one calls this evolution a backward evolution, and it
has previously been used~\cite{Mueller:2003bz} in studying shadowing
effects in scattering on nuclei. The Kovchegov equation in this
backward evolution is imposed approximately in terms of the boundary
$\rho_2$. There is in fact a complete symmetry here; the boundary
$\rho_1$ is to $\rho_i$ in forward evolution as the boundary $\rho_2$
is to $\rho_f$ in backward evolution (see Fig.~\ref{Fig_int_2b}).

The introduction of the second boundary $\rho_2$ and the resulting
symmetry which it brings with it has another benefit. The scattering
amplitude is now boost invariant which was not the case for the
single boundary case. Thus the calculation in the center of mass, say, now
gives the same answer as arises when all the evolution is put into one
or the other of the two dipoles.

The paper is organized as follows: Sec.~\ref{Sec_DD} gives some
results from dipole-dipole scattering in the BFKL approximations. In
Sec.~\ref{Sec_The Kovchegov_Equation_and_its_Limitations} the
Kovchegov equation is derived emphasizing the ``mean field'' aspects
necessary for the derivation. The approximate solution is given in the
absorptive boundary approximation. The frame-dependence of scattering
is demonstrated. In Sec.~\ref{BFKL_DW_SOL} the second boundary,
$\rho_2$, is introduced and the BFKL equation is solved in the
presence of the two boundaries in the diffusion approximation. The
positions of the boundaries and the energy dependence of the
saturation momentum are determined. In Sec.~\ref{Sec_Running_Coupling}
the discussion of Sec.~\ref{BFKL_DW_SOL} is repeated for running
coupling evolution.

In Sec.~\ref{BFKL_DW_SOL} and Sec.~\ref{Sec_Running_Coupling} we have
solved the BFKL equations, with boundaries $\rho_1$ and $\rho_2$, in a
diffusion approximation. Perhaps the main change from the single
boundary case is that the scaling law which follows from the Kovchegov
equation is now badly violated as explained at the end of
Sec.~\ref{Sec_BFKL_dw}. The effect of the two boundaries, as compared
to the single boundary, is parametrically small but in practice very
large. For example the anomalous dimension $\lambda_0$ becomes
$\lambda_d$ where
\be
\lambda_d = \lambda_0 + \frac{1}{2(1-\lambda_0)}
\frac{\pi^2}{(\rho_2-\rho_1)^2}
\ee
and where $\rho_2-\rho_1 \approx \ln1/\alpha^2 + \rho_f-\rho_{is}(Y)$
($\rho_{is}$ is the saturation line for internal evolution),
parametrically. Indeed the resulting corrections are so large that one
may well doubt the diffusion approximation we have used. It's possible
that our procedure has overcorrected the deficiencies of the Kovchegov
equation, although our second boundary only eliminates paths of
evolution which manifestly violate unitarity. It would be interesting
to do a numerical solution of BFKL evolution in the presence of two
boundaries to see if the diffusion approximation is accurate.

We feel that the main conclusion of this study is that correlations may
be very important near the saturation boundary and in the scaling
region. The best way to see if this is really true, and to see if our
procedure of calculation has hit the essence of the physics, is to do
accurate numerical simulations of the JIMWLK equation as already begun
in Ref.~\cite{Rummukainen:2003ns} to see how different those results
are from those of the Kovchegov equation and from our two boundary
procedure of imposing unitarity.

\section{Dipole-Dipole scattering; general definitions}
\label{Sec_DD}
In this section we give some general formulas and definitions concerning
the dipole-dipole scattering which will be used throughout this work.

Suppose we consider the scattering of a dipole of size $x$ on a dipole
of size $x'$ at relative rapidity $Y$. In a frame where one of the
dipoles has rapidity $y$ and the other has rapidity $Y-y$, the forward
scattering amplitude is given by
\be
T(x,x',Y) = 
        \int \frac{d^2r_1 d^2r_2}{4\pi^2 r_1^2 r_2^2}\ n(x,r_1,y)
        \ n(x',r_2,Y-y)\ \sigma_{dd}(r_1,r_2) 
\label{eq_T_cm}
\ee
where $\sigma_{dd}(r_1,r_2)$ is the
dipole-dipole cross section at the two gluon exchange level 
\be
\sigma_{dd}(r_1,r_2) =
         2 \pi \alpha^2 r_<^2 \left(1 + \ln\frac{r_>}{r_<}\right) \ ,
\label{eq_sig_dd}
\ee
with $r_<$ being the smaller of $r_1$, $r_2$ and $r_>$ being the
larger of $r_1$, $r_2$ while $n(x,r_1,y)$ is the number density of
radiated dipoles of size $r_1$ in the wavefunction of a parent dipole
of size $x$ in the rapidity interval $y$. The expression for
$n(x,r_1,y)$ which obeys the dipole version of the BFKL equation reads
for $x > r_1$~\cite{Mueller:1999yb}
\be
n(x,r_1,y) = 2 \left(\frac{x^2}{r_1^2}\right) \int 
            \frac{d\lambda}{2\pi i}\ 
             \exp\!\left[\frac{2\alpha
             N_c}{\pi}{\chi}(\lambda) y - (1-\lambda)\ln\frac{x^2}{r_1^2}\right]  
\label{eq_n_ls}
\ee
and for $x < r_1$ 
\be
n(x,r_1,y) = 2 \int \frac{d\lambda}{2 \pi i}\ \exp\!\left[\frac{2\alpha
             N_c}{\pi}{\chi}(\lambda) y - (1-\lambda)\ln\frac{r_1^2}{x^2}\right]  
\label{eq_n_sl}
\ee
or
\be
n(x,r_1,y) = \left(\frac{x^2}{r_1^2}\right) n(r_1,x,y) \quad
(\mbox{for $x < r_1$)} \ ,
\label{eq_rel_nls}
\ee
where
\be
\chi(\lambda) = \psi(1) -\frac{1}{2} \psi(\lambda) - \frac{1}{2}
                 \psi(1-\lambda)
\label{eq_chi}
\ee
with $\psi(\lambda)=\Gamma'(\lambda)/ \Gamma(\lambda)$. The
integration contour in eqs.~(\ref{eq_n_ls}) and (\ref{eq_n_sl}) is
parallel to the imaginary axis with $0 < \mbox{Re}(\lambda) < 1$.

The scattering amplitude in eq.~(\ref{eq_T_cm}) can be evaluated in an
arbitrary frame since the result does not depend on the frame. Often
the evalution of the amplitude becomes especially easy in the
laboratory frame where one of the dipoles carries all of the rapidity
(this dipole evolves) and the other one has zero rapidity (this dipole
is an elementary dipole). In this case, for $y=Y$, the integration
over $r_2$ can be done because $n(x',r_2,0) = r_2 \delta(r_2-x')$
and the result is
\be
T(x,x',Y) = 
        \int \frac{d^2r_1}{2\pi r_1^2}\ n(x,r_1,Y)\
        \sigma_{dd}(r_1,x') \ . 
\label{eq_T_lab}
\ee
Inserting (\ref{eq_sig_dd}) and (\ref{eq_n_ls}) in
(\ref{eq_T_lab}), also the integration over $r_1$ can be easily done, 
\be
\int \frac{d^2r_1}{2 \pi r_1^2}
  \left(\frac{x^2}{r_1^2}\right)^{\lambda} 
   \ 2 \pi \alpha^2 r_<^2 \left(1 + \ln\frac{r_>}{r_<}\right)
  = \frac{\pi}{2} \alpha^2 x'^2 \left(\frac{x^2}{x'^2}\right)^{\lambda}
  \frac{1}{\lambda^2 (1-\lambda)^2} \ ,
\label{eq_lam}
\ee
leading to  
\be
T(x,x',Y) = 
        \pi \alpha^2 x'^2 \int \frac{d \lambda}{2\pi i}  
        \frac{1}{\lambda^2 (1-\lambda)^2} \exp\!\left[\frac{2\alpha
             N_c}{\pi}{\chi}(\lambda) Y -
           (1-\lambda)\ln\frac{x^2}{x'^2}\right] 
          \left(\frac{x^2}{x'^2}\right) \ .
\label{eq_T_lab_lam}
\ee
%

%__________The Kovchegov Equation and its Limitations___________________________
\section{The Kovchegov equation and its limitations}
\label{Sec_The Kovchegov_Equation_and_its_Limitations}
% ______________________________________________________________________________
In this section we ``derive'' the Kovchegov
equation~\cite{Kovchegov:1999yj+X} which is probably the best
``simple'' equation including nonlinear evolution in QCD. An analytic
solution to the Kovchegov equation in the region close to the
saturation regime has been recently achieved~\cite{Munier:2003vc+X};
analytic results for the scattering amplitude in laboratory frame and
for the saturation momentum as a function of rapidity are obtained.
These results agree with the ones obtained before by solving the
linear BFKL equation in the presence of a saturation
boundary~\cite{Mueller:2002zm}. In this work, we show that the
scattering amplitude in the vicinity of the saturation regime which
results from the Kovchegov equation does not satisfy the completeness
relation and it is frame-dependent.

%leads to inconsistences if the scattering proccess is
%considered in different frames. 
%The Kovchegov equation is of limited
%validity since it is a mean field approximation of a more complete
%equation.

%-----------------------------------------------------------
\subsection{The Kovchegov equation}
\label{Sec_Kovchegov_equation}
\begin{figure}[h!]
\setlength{\unitlength}{1.cm}
\begin{center}
\epsfig{file=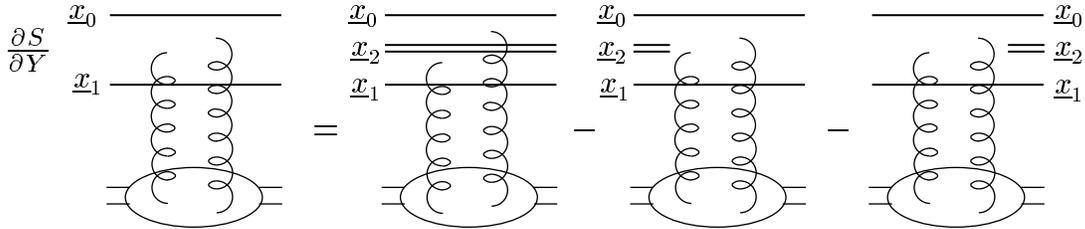, width=14.5cm}
\end{center}
\caption{Graphs corresponding to terms in the Kovchegov equation.}
\label{Fig_Kovchegov}
\end{figure}

%------------------------------------------------------------
Consider the high-energy scattering of a quark-antiquark dipole on a
target which may be another dipole, a hadron or a nucleus.  For
convenience we view the scattering process in a frame where the dipole
is left moving and the target is right moving. Further we suppose that
almost all of the relative rapidity of the dipole and the target,
$Y$, is taken by the target so that the probability for the
left-moving dipole to emit gluons before scattering off the target is
small. In this frame an elementary quark-antiquark dipole scatters on
a highly evolved target.  We denote the corresponding elastic
scattering amplitude by $S({\underline x}_0,{\underline x}_1,Y)$ where
${\underline x}_0$ and ${\underline x}_1$ are the transverse
coordinates of the quark and antiquark of the dipole, respectivelly.
Now we wish to know how $S(\underline{x}_0,\underline{x}_1,Y)$ changes
when the rapidity $Y$ is increased by a small amount $dY$. The
increase $dY$ can be viewed either as increasing the momentum of the
target or as increasing the momentum of the elementary dipole. If the
rapidity of the target is increased than its wavefunction evolves
further. This evolution is given by the functional equation derived by
Jalilian-Marian, Iancu, McLerran, Leonidov and Kovner
(JIMWLK)~\cite{Iancu:2001ad+X,Jalilian-Marian:1997jx+X,Weigert:2000gi}.
On the other hand, if the rapidity of the dipole is increased while
that of the target is kept fixed, the probability for the dipole to
emit a gluon increases (proportional to $dY$) due to the change $dY$.
In the large $N_c$ limit this quark-antiquark-gluon state can be
viewed as a system of two dipoles -- one of the dipoles consists of
the initial quark and the antiquark part of the gluon while the other
dipole is given by the quark part of the gluon and the initial
antiquark. The probability for the production of a
quark-antiquark-gluon state from the initial quark-antiquark dipole
is~\cite{Mueller:1994rr}
\be
dP = \frac{\alpha N_c}{2 \pi^2} d\underline{x}_2 dY 
     \frac{\underline{x}^2_{01}}{\underline{x}^2_{02}
       \underline{x}^2_{12}} \ ,
\ee
where $\underline{x}_2$ is the transverse coordinate of the emitted
gluon while $\underline{x}_{01}$, $\underline{x}_{02}$ and
$\underline{x}_{12}$ with
$\underline{x}_{ij}=\underline{x}_i-\underline{x}_j$ are the
transverse sizes of the dipoles as shown in Fig.~\ref{Fig_Kovchegov}.
The probability $dP$ when multiplied with the $S$-matrix for the two
dipole state to elastically scatter on the target gives the change in
the $S$-matrix, $dS$, for dipole-hadron scattering
\be 
\frac{\partial}{\partial Y} S(\underline{x}_{01},Y) =
           \frac{\alpha N_c}{2 \pi^2} \int d^2{\underline x}_2
           \frac{\underline{x}^2_{01}}{\underline{x}^2_{02} 
           \underline{x}^2_{12}} \left [
           S^{(2)}(\underline{x}_{02},\underline{x}_{12},Y) -
           S(\underline{x}_{01},Y) \right ] \ .
\label{eq_kov_1}
\ee
Here $S^{(2)}(\underline{x}_{02},\underline{x}_{12},Y)$ stands for the
scattering of the two dipole state on the target (first graph on the
left-hand side of Fig.~\ref{Fig_Kovchegov}) while the subtracted
$S(\underline{x}_{01},Y)$ is the virtual contribution necessary to
normalize the wavefunction~\cite{Mueller:1994rr,Munier:2003zb}. The
later gives the scattering of a single dipole on the target because
the gluon is not in the wavefunction of the dipole at the time of the
scattering (last two graphs in Fig.~\ref{Fig_Kovchegov}).

Eq.~(\ref{eq_kov_1}) is part of a set of equations derived by
Balitsky~\cite{Balitsky:1995ub+X} and it also results from the JIMWLK
functional equations~\cite{Iancu:2001ad+X,Weigert:2000gi}. It is difficult to use
eq.~(\ref{eq_kov_1}) because of the unknown
$S^{(2)}(\underline{x}_{02},\underline{x}_{12},Y)$. The assumption
that the scattering of the two dipole state on the target factorizes
\be
S^{(2)}(\underline{x}_{02},\underline{x}_{12},Y) =
  S(\underline{x}_{02},Y) S(\underline{x}_{12},Y) \ ,
\label{fac}
\ee
which is a sort of a mean field approximation for the gluonic fields
in the target, leads to the Kovchegov equation~\cite{Kovchegov:1999yj+X}
\be 
\frac{\partial}{\partial Y} S(\underline{x}_{01},Y) =
           \frac{\alpha N_c}{2 \pi^2} \int d^2{\underline x}_2
           \frac{\underline{x}^2_{01}}{\underline{x}^2_{02} 
           \underline{x}^2_{12}} \left [
           S(\underline{x}_{02},Y) S(\underline{x}_{12},Y) -
           S(\underline{x}_{01},Y) \right ] \ .
\label{Eq_Kovchegov}
\ee
This equation is probably the best ``simple'' equation taking into
account unitarity corrections.  In his original
works~\cite{Kovchegov:1999yj+X} Kovchegov used the
approximation~(\ref{fac}) for the case that the target is a large
nucleus. In this case, eq.~(\ref{fac}) is a reasonable approximation.
The factorization~(\ref{fac}) is however less obvious for an arbitrary
target. It is the mean field approximation (or factorization)
which makes the Kovchegov equation treat fluctuations incorrectly.
In the region where $S$ is small, it has been shown in
Ref.~\cite{Iancu:2003zr} that the Kovchegov equation, while giving the
form of the $S$-matrix correctly, gives the exponential factor twice
as large as the result which emerges when flutuations are taken into
account. Furthermore, in the next subsections, we show that the
scattering amplitude which results from the Kovchegov equation is not
consistent with the completeness relation and it is frame-dependent.
This is because the evolution of the amplitude in rapidity is devided
into two successive steps in both cases and at the intermediate step
fluctuations are not properly taken into account.

Using ${\cal N}(\underline{x}_{ij},Y) = 1-S(\underline{x}_{ij},Y)$, 
another useful version of the Kovchegov equation follows
\bea 
\frac{\partial}{\partial Y} N(\underline{x}_{01},Y) =
           \frac{\alpha N_c}{2 \pi^2} \int d^2{\underline x}_2
           \frac{\underline{x}^2_{01}}{\underline{x}^2_{02} 
           \underline{x}^2_{12}}\!\!\!\!&&\!\!\!\left [
           N(\underline{x}_{02},Y) + N(\underline{x}_{12},Y) \right .
           \nonumber \\
          && \left . \!\!\!-N(\underline{x}_{01},Y)- N(\underline{x}_{02},Y)
           N(\underline{x}_{12},Y) \right ] \ .
\label{Eq_Kovchegov_N}
\eea
This equation becomes easy to use when the scattering is weak since
the nonlinear term $N(\underline{x}_{02},Y)N(\underline{x}_{12},Y)$
can be dropped and the linear equation remaining is the dipole
version~\cite{Mueller:1994rr} of the Balitsky-Fadin-Kuraev-Lipatov
(BFKL) equation~\cite{Kuraev:fs,Balitsky:ic} whose solution is known
in the saddle-point approximation~\cite{Mueller:1994rr}. In the
high-energy regime where unitarity corrections become important or
$S(\underline{x}_{ij},Y)$ is small, eq.~(\ref{Eq_Kovchegov}) is easier
to use since the quadratic term $S(\underline{x}_{02},Y)
S(\underline{x}_{12},Y)$ can be neglected.  The solution in this
region is also known~\cite{Levin:1999mw,Mueller:2001fv}. Also the
transition region from the weak to the saturation regime (sometimes
also called the ``scaling region'') has been explored; analytic
expressions for the rapidity dependence of the saturation momentum and
for the scattering amplitude have been extracted by the authors of
Ref.~\cite{Mueller:2002zm} by solving the BFKL equation in the
presence of a saturation boundary.  These results have been recently
confirmed by Munier and Peschanki~\cite{Munier:2003vc+X} who have
directly solved the Kovchegov equation in the scaling region. In the
next section, we show these results and discuss their validity.

%As we will see in the next
%section, this result is inconsistent with the completeness relation
%and considerations in laboratory versus center of mass (c.m.) frame.

%--------------------------------------------------------------
\subsection{Solution to the Kovchegov equation}
\label{Sec_Solution_Kovchegov}
%--------------------------------------------------------------

\begin{figure}[h!]
\setlength{\unitlength}{1.cm}
\begin{center}
\epsfig{file=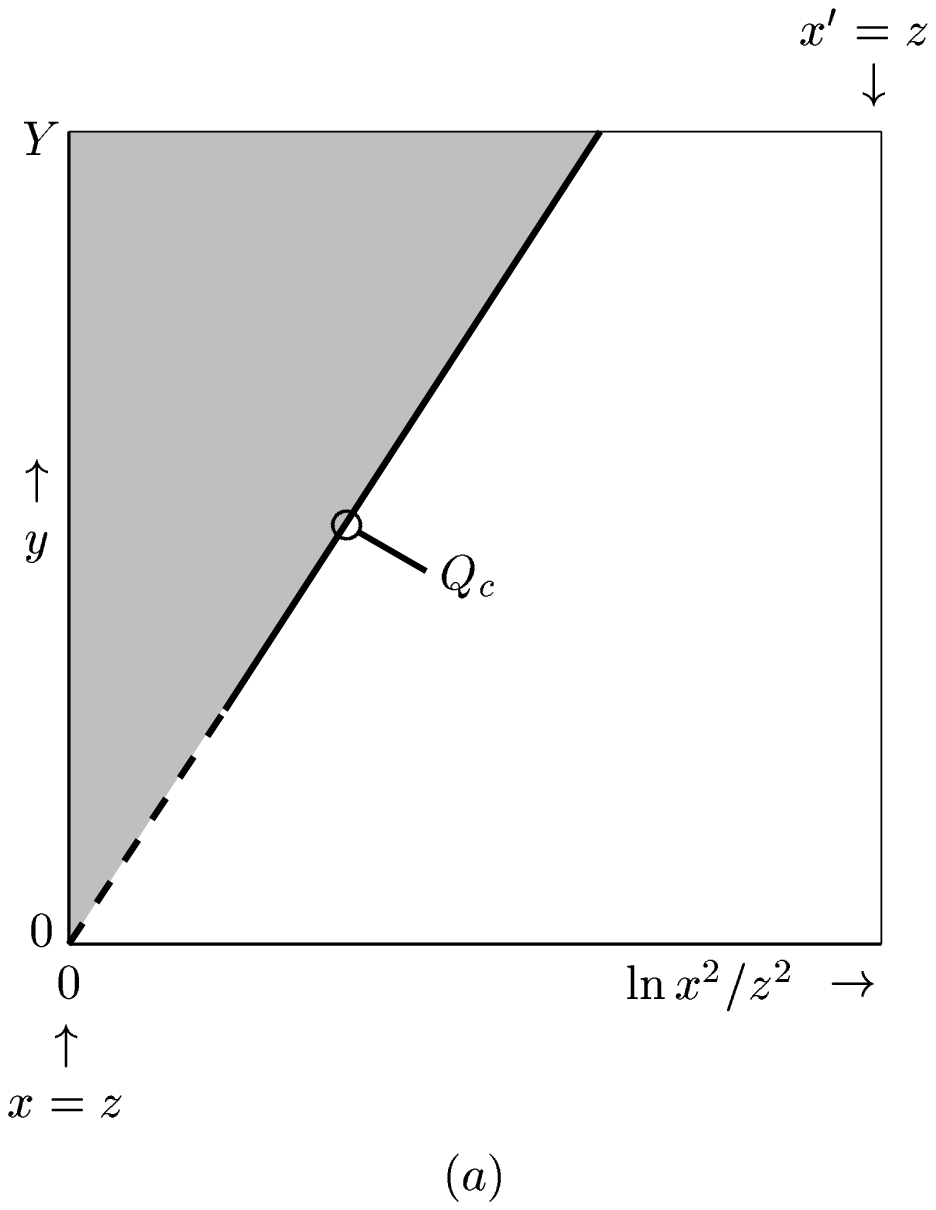, width=6.5cm}
\hfill
\epsfig{file=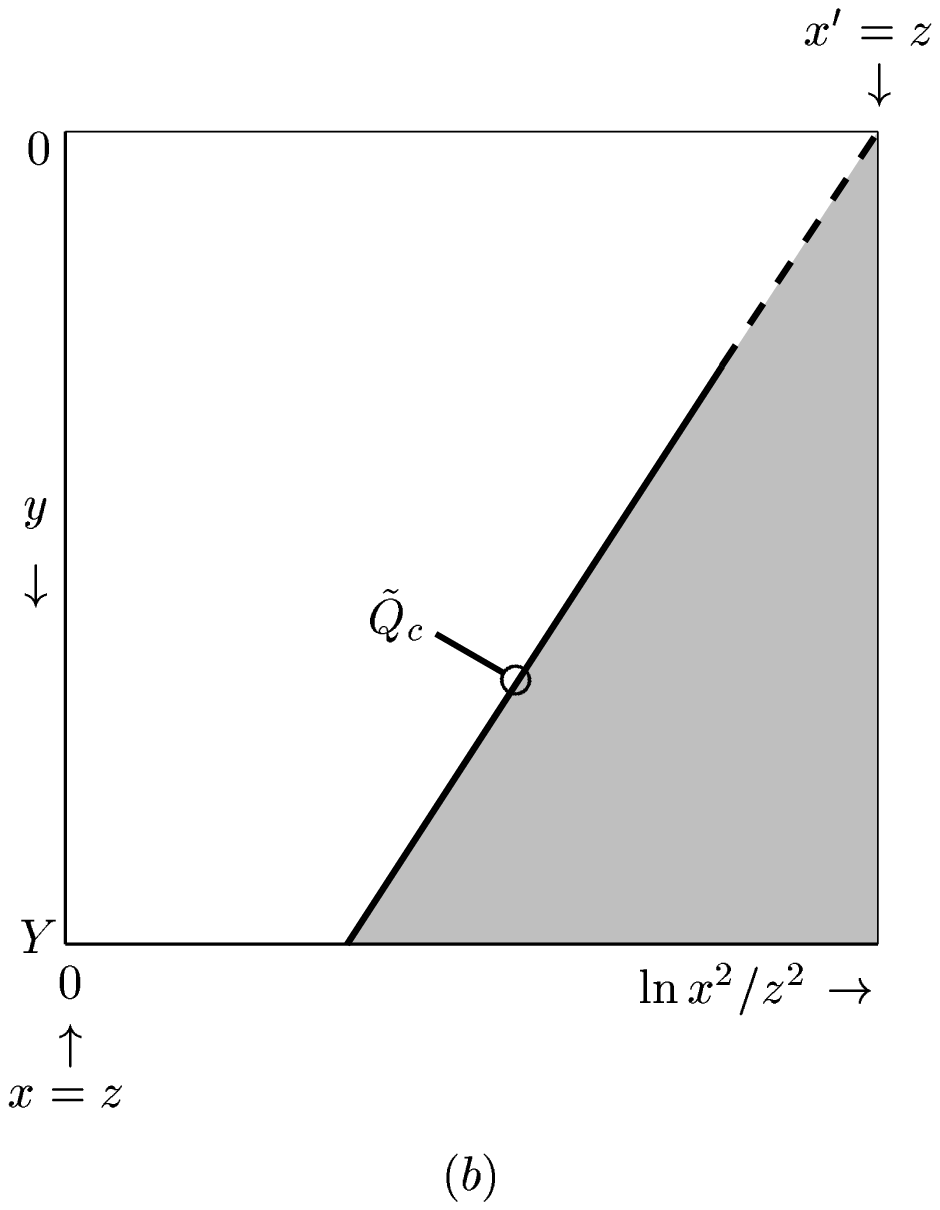, width=6.5cm}
\end{center}
\caption{The saturation
  boundaries $Q_c$ and $\tilde{Q}_c$ separate the saturated regime
  (shaded areas) from the weak interaction region. In (a) dipole $x$
  evolves from $y=0$ up to $y=Y$ and scatters on the elementary dipole
  $x'$ while in (b) dipole $x'$ evolves from $y=0$ up to $y=Y$ and
  scatters on the elementary dipole $x$.}
\label{Fig_T_ls}
\end{figure}

In the vicinity of the saturation regime, the amplitude for
dipole-dipole scattering in laboratory frame emerging from the
Kovchegov equation~\cite{Munier:2003vc+X} or, equivalently, from the
BFKL evolution in the presence of a saturation
boundary~\cite{Mueller:2002zm}, is\footnote{The scattering amplitude
  shown here follows from eq.(4.1) of Ref.~\cite{Mueller:2002zm} in
  the vicinity of the saturation boundary, including the exponential
  function, which was neglected in eq.(4.2) there.}
\bea
\!\!\!\!\!T(x,x',Y) = \frac{4\sqrt{\pi} \alpha^2 x^2}{\lambda_0^2 (1-\lambda_0)^2} 
            \left[Q_c^2(Y)\,x'^2\right]^{1-\lambda_0}
            &&\!\!\!\!\!\!\!\!\!\!\Delta\left[\ln\frac{1}{Q^2_c(Y)\,x'^2} + \Delta\right]
            \nonumber \\
            &&\!\!\!\!\!\!\!\!\!\!\times
            \exp\!\left[-\frac{\pi
                \ln^2(1/Q^2_c(Y)\,x'^2)}{4 \alpha N_c
                \chi''(\lambda_0)Y}\right] \ . 
\label{eq_T_fQ}
\eea
Here all the values for the constants and the notations are adopted
from Ref.~\cite{Mueller:2002zm}. The above amplitude results from the
scattering amplitude (\ref{eq_T_lab_lam}) when evaluated in the
vicinity of the boundary line $Q_c(Y)$, as done
in~\cite{Mueller:2002zm}, where
\be
Q^2_c(Y)\ x^2 = \frac{\exp\!\left[\frac{2\alpha
      N_c}{\pi}\frac{\chi(\lambda_0)}{1-\lambda_0}Y\right]}
     {\left[\frac{4\alpha
           N_c}{\pi}\chi''(\lambda_0)Y\right]^{\frac{3}{2
           (1-\lambda_0)}}} 
\label{eq_Q_c}
\ee
with $\lambda_0 = 0.372$ determined by
\be
(1-\lambda_0) \chi'(\lambda_0) + \chi(\lambda_0) = 0 \ . 
\label{eq_lambda_0}
\ee
Up to a constant factor, $Q_c(Y)$ equals the saturation momentum $Q_s(Y)$, 
which separates the saturation regime from the weak scattering region.
In Fig.\ref{Fig_T_ls}a we show the line $Q_c(Y)$ in the
$\ln{x^2/z^2}-Y$ plane where $z$ is generic; BFKL evolution takes
place on the right-hand side of the line. All paths going into the
saturation region on the left-hand side of the line are excluded.

The scattering amplitude (\ref{eq_T_fQ}) is valid in the region
\be
 \Delta \leq \ln\frac{1}{Q_c^2(Y)\, x'^2} \ll \sqrt{4 \alpha N_c
\chi''(\lambda_0) Y/\pi}  \ ,
\label{eq_T_val}
\ee
and $\Delta$ is an irrelevant small constant depending on the
fixed coupling $\alpha$.  

In the vicinity of the saturation boundary, i.e., for $x'^2$ close to
$1/Q^2_c(Y)$, the scattering amplitude in (\ref{eq_T_fQ}) shows a
scaling behaviour since it depends only on $Q_c^2(y)\,x'^2$; lines
with constant $Q_c^2(y)\,x'^2$ are lines of constant scattering
amplitude.

Expression (\ref{eq_T_fQ}) can be viewed as follows; one takes a
dipole of size $x$ at zero rapidity, evolves it to many smaller
dipoles (their size is of order $1/Q_c(Y)$) as the rapidity increases
up to $Y$, and scatters then these dipoles on an elementary dipole of
size $x'$. This is shown in Fig.~\ref{Fig_T_ls}a. Equivalently, one may
view the BFKL evolution also in the opposite direction, as shown in
Fig.~\ref{Fig_T_ls}b, in which case the small dipole $x'$ evolves to larger
dipoles (size of order $1/\tilde{Q}_c(Y)$) which then scatter on the
elementary dipole $x$. In this case the
scattering amplitude takes the form
\bea 
\!\!\!\!\!T(x',x,Y) = \frac{4\sqrt{\pi} \alpha^2 x^2}{\lambda_0^2
                   (1-\lambda_0)^2}
                 \left[\frac{1}{\tilde{Q}_c^2(Y)\,x^2}\right]^{1-\lambda_0}
&&\!\!\!\!\!\!\!\!\!\!\Delta\left[\ln\tilde{Q}^2_c(Y)\,x^2 +
  \Delta\right]
\nonumber \\
&&\!\!\!\!\!\!\!\!\!\!\times \exp\!\left[-\frac{\pi
    \ln^2(\tilde{Q}^2_c(Y)\,x^2)}{4 \alpha N_c
    \chi''(\lambda_0)Y}\right] \ ,
\label{eq_T_fQ_2}
\eea
where $\tilde{Q}_c^2(Y)$ is defined by
\be
 Q^2_c(Y)\,x'^2 = \frac{1}{\tilde{Q}^2_c(Y)\,x^2} \ ,
\label{eq_Q_Qt}
\ee
which is obvious from Figs.~\ref{Fig_T_ls}a and \ref{Fig_T_ls}b. This
additional way of viewing the scattering of two dipoles turns out to
be useful in our further discussion, as we will see. 

It is also useful to show the expressions for the dipole number
densities in the vicinity of the saturation boundary. They are
calcuated in the same way as the scattering amplitudes.  The dipole
number density for $x>x'$ reads
\be
n(x,x',Y)\!=\!\frac{8}{\sqrt{\pi}}\!\left(\!\frac{x^2}{x'^2}\!\right)\!\!\left[Q_c^2(Y)\,x'^2\right]^{1-\lambda_0}
            \!\Delta\!\!\left[\ln\frac{1}{Q^2_c(Y)\,x'^2}\!+\!\!\Delta\right]
             \!\exp\!\left[\!-\frac{\pi
                \ln^2(1/Q^2_c(Y)x'^2)}{4 \alpha N_c
                \chi''(\lambda_0)Y}\!\right]
\label{eq_n_sat}
\ee
and it is related to the scattering amplitude as
\be
T(x,x',Y) = \frac{\pi}{2} \alpha^2 x'^2 \frac{1}{\lambda_0^2
  (1-\lambda_0)^2}\ n(x,x',Y) \ ,
\label{eq_T_n}
\ee
which is obvious from eqs.~(\ref{eq_n_ls}), (\ref{eq_T_lab_lam}) and
(\ref{eq_T_fQ}). Eq.~(\ref{eq_T_n}) explicitly shows that the
elementary dipole $x'$ scatters on the evolved dipoles $x$.

Similary, for the evolution from $x'$ to $x$ (with $x'<x$), one obtains
\be 
n(x',x,Y) = \frac{8}{\sqrt{\pi}}\left[\frac{1}{\tilde{Q}_c^2(Y)\,x^2}\right]^{1-\lambda_0}
            \Delta\left[\ln\tilde{Q}^2_c(Y)\,x^2 +\Delta\right]
            \exp\!\left[-\frac{\pi \ln^2(\tilde{Q}^2_c(Y)\,x^2)}{4 \alpha N_c
    \chi''(\lambda_0)Y}\right] \ ,
\label{eq_n_sl_sat}
\ee
and has the relation
\be
T(x',x,Y) = \frac{\pi}{2} \alpha^2 x^2 \frac{1}{\lambda_0^2
  (1-\lambda_0)^2}\ n(x',x,Y) \ ,
\label{eq_T_n_sl}
\ee
as can be seen from (\ref{eq_n_sl}), (\ref{eq_T_lab_lam}) and
(\ref{eq_T_fQ_2}). As compared to (\ref{eq_T_n}), eq.~(\ref{eq_T_n_sl}) shows
that the elementary dipole $x$ scatters on the evolved dipole $x'$.

\subsection{Completeness relation}
\subsubsection{Internal saturation boundaries}
\begin{figure}[h!]
\setlength{\unitlength}{1.cm}
\begin{center}
\epsfig{file=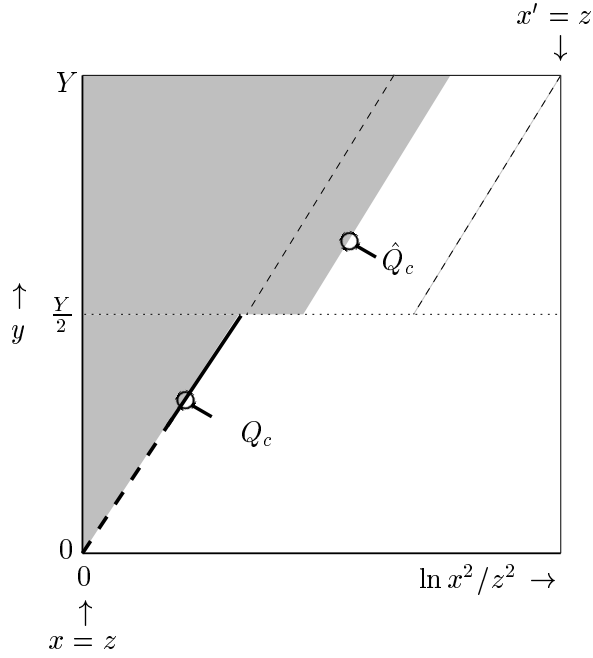, width=7.5cm}
\end{center}
\caption{BFKL evolution in two successive steps $0 \to Y/2 \to Y$ in
  the presence of the internal saturation boundaries $Q_c$ and $\hat{Q}_c$.}
\label{Fig_BFKL_CR}
\end{figure}

In this section we show that the scattering amplitude in the laboratory
frame~(\ref{eq_T_fQ}) resulting from the Kovchegov equation (or the
BFKL evolution with a saturation boundary) is not consistent with the
completeness relation. This becomes obvious if one tries to reproduce
the result (\ref{eq_T_fQ}) by dividing the BFKL evolution in two
successive steps in rapidity, say from $y=0$ to $y=Y/2$ and then from
$y=Y/2$ to $y=Y$ as shown in Fig.~\ref{Fig_BFKL_CR}. To do this, we
start with the scattering amplitude of two dipoles in the
center-of-mass frame which is given by (\ref{eq_T_cm}) with $y=Y/2$
\be
T(x,x',Y) = 
        \int \frac{d^2r_1 d^2r_2}{4\pi^2 r_1^2 r_2^2}\ n(x,r_1,Y/2)
        \ n(x',r_2,Y/2)\ \sigma_{dd}(r_1,r_2) \ . 
\label{eq_T_cm_1}
\ee
Upon inserting the dipole number density (\ref{eq_n_ls}), where $x > r_1$,
and the dipole-dipole cross section (\ref{eq_sig_dd}) in
(\ref{eq_T_cm_1}), the integration over $r_1$ can be carried out as in
(\ref{eq_lam}), and the amplitude becomes
\bea
T(x,x',Y) = 
        \pi \alpha^2 \!\!\int\!\!\frac{d^2r_2}{2 \pi} \int\!\!\!\frac{d \lambda}{2\pi i}  
         &&\!\!\!\!\!\!\!\!\!        
\frac{1}{\lambda^2 (1-\lambda)^2} \exp\!\left[\frac{2\alpha
             N_c}{\pi}{\chi}(\lambda) \frac{Y}{2} -
           (1-\lambda)\ln\frac{x^2}{r_2^2}\right]\!\! 
          \left(\frac{x^2}{r_2^2}\right) 
          \nonumber\\
          && \!\!\!\!\!\!\!\times \ n(x',r_2,Y/2) \ .
\label{eq_T_lab_lam_0}
\eea
Using relation (\ref{eq_rel_nls}) for $x'<r_2$, one can rewrite the amplitude as 
\bea
T(x,x',Y) = 
        \pi \alpha^2 x'^2\!\!\int\!\!\frac{d^2r_2}{2 \pi r_2^2} \int\!\!\!\frac{d \lambda}{2\pi i}  
         &&\!\!\!\!\!\!\!\!\!        
\frac{1}{\lambda^2 (1-\lambda)^2} \exp\!\left[\frac{2\alpha
             N_c}{\pi}{\chi}(\lambda) \frac{Y}{2} -
           (1-\lambda)\ln\frac{x^2}{r_2^2}\right]\!\! 
          \left(\frac{x^2}{r_2^2}\right) 
          \nonumber\\
          && \!\!\!\!\!\!\!\times \ n(r_2,x',Y/2) 
\label{eq_T_lab_lam_1}
\eea
which further can be expressed in terms of dipole number densities (cf.
eq.~(\ref{eq_n_ls})),
\be
T(x,x',Y) = \frac{\pi}{2} \alpha^2 x'^2 \frac{1}{\lambda_0^2
    (1-\lambda_0)^2} \int\!\!\frac{d^2r_2}{2 \pi r_2^2}\ n(x,r_2,Y/2)\ 
    n(r_2,x',Y/2) \ .
\label{eq_T_lab_lam_2}
\ee
It is at this point that we can identify the completeness relation.  
Requiring this expression to be the same as the result for the
scattering amplitude in the laboratory frame (\ref{eq_T_fQ}),  
or equivalently (\ref{eq_T_fQ_2}), means that the following
completeness relation, 
\be
n(x,x',Y) = \int\!\!\frac{d^2r_2}{2 \pi r_2^2}\ n(x,r_2,Y/2)\ 
n(r_2,x',Y/2) \ ,
\label{eq_cr}
\ee 
has to be satisfied for the dipole number densities $n$ evaluated in
the presence of saturation boundaries. We note in passing that one can
use eq.~(\ref{eq_n_ls}) to show that the naive BFKL evolution without
saturation boundaries fulfilles the completeness relation.

Let us now check if the right-hand side (rhs) of (\ref{eq_cr}) equals
the left-hand side which is given by eq.~(\ref{eq_n_sat}). Inserting
the dipole number density (\ref{eq_n_sat}) with $Y \to Y/2$ in
(\ref{eq_cr}), the right-hand side becomes
\bea
\mbox{[rhs of (\ref{eq_cr})]} &=& \frac{64}{\pi}\int \frac{d^2r_2}{2\pi r_2^2} 
\label{eq_rhs_cr} \\ 
                 &&\hspace{-3cm}\times \left(\frac{x^2}{r_2^2}\right) 
                 \left[Q_c^2(Y/2)\,r_2^2\right]^{1-\lambda_0}
             \Delta \left[\ln\frac{1}{Q^2_c(Y/2)\,r_2^2} + \Delta\right]
             \exp\!\left[\!-\frac{\pi
                \ln^2(1/Q^2_c(Y/2)r_2^2)}{4 \alpha N_c
                \chi''(\lambda_0)Y/2} \right]
                \nonumber\\
                && \hspace{-3cm}\times \left( \frac{r_2^2}{x'^2}\right) 
             \left[\hat{Q}_c^2(Y/2)\,x'^2\right]^{1-\lambda_0}
             \Delta \left[\ln\frac{1}{\hat{Q}^2_c(Y/2)\,x'^2} + \Delta\right]
              \exp\!\left[\!-\frac{\pi
                \ln^2(1/\hat{Q}^2_c(Y/2)x'^2)}{4 \alpha N_c
                \chi''(\lambda_0)Y/2} \right],\nonumber
\eea
where $Q^2_c(Y/2)$ follows from (\ref{eq_Q_c}) with $Y$ replaced by
$Y/2$, and $\hat{Q}^2_c(Y/2)$ is determined by matching the evolution
at $Y/2$ (see Fig.~\ref{Fig_BFKL_CR}),
\be
\hat{Q}_c^2(Y/2)\,r_2^2 = Q_c^2(Y/2)\,x^2 \ .
\label{eq_matching}
\ee
In equation (\ref{eq_rhs_cr}) the second line gives the evolution of
dipole $x$ from $y=0$ up to $y=Y/2$ and the third line gives the
evolution of dipole $r_2$ from $Y/2$ up to $Y$. It is important to
note that while the evolution of dipole $x$ is limited by the
saturation boundary $Q_c$, the evolution of dipole $r_2$ is limited by
its {\em own} (or {\em internal}) saturation boundary $\hat{Q}_c$ as
shown by the two shaded areas in Fig.~\ref{Fig_BFKL_CR} in order to
respect unitarity.

Using the matching
(\ref{eq_matching}) which gives
\bea
\!\!\!\!\!\!\!\!\!\left[Q_c^2(Y/2)\,r_2^2\right]^{1-\lambda_0}
\left[\hat{Q}_c^2(Y/2)\,x'^2\right]^{1-\lambda_0} &=&\!\!\left[Q^4_c(Y/2)
  x^2 x'^2 \right]^{1-\lambda_0} \nonumber \\
 &=&\!\!\left[Q^2_c(Y) x'^2\right]^{1-\lambda_0}
 \frac{1}{\left[\frac{\alpha N_c}{\pi} \chi''(\lambda_0)
     Y\right]^{\frac{3}{2}}} \ ,
\label{eq_def_z_e_t}
\eea
and the definitions 
\be
\eta = \ln\frac{1}{Q^2_c(Y/2)\,r_2^2} \ , \quad 
\zeta = \ln\frac{1}{Q^4_c(Y/2) x^2 x'^2} \ , \quad 
\tau = \frac{\alpha N_c}{\pi} \chi''(\lambda_0) Y  \ ,
\ee
equation (\ref{eq_rhs_cr}) becomes

\bea
\mbox{[rhs of (\ref{eq_cr})]} &=& \frac{8}{\pi}\left(\frac{x^2}{x'^2}\right) 
                  \left[Q_c^2(Y)\,x'^2\right]^{1-\lambda_0} 
                    \frac{1}{\tau^{\frac{3}{2}}}
                  \int_{-\Delta}^{\zeta+\Delta} d\eta 
\label{eq_rhs_cr_2}
                 \\ 
                 && \times  \Delta \left[\eta + \Delta \right]
                 \exp\!\left[-\frac{\eta^2}{2 \tau}\right]
                 \Delta \left[\zeta - \eta + \Delta \right]
                \exp\!\left[-\frac{(\zeta-\eta)^2}{2 \tau} \right] \ .\nonumber
\eea
The limits in the intgration over $\eta$ follow from the positive
definite dipole number densities (\ref{eq_T_val}). The integration
over $\eta$ is easily done and the result is
\bea
&&\hspace{-0.6cm}\mbox{[rhs of (\ref{eq_cr})]} = \frac{8}{\pi}\left(\frac{x^2}{x'^2}\right) 
                  \left[Q_c^2(Y)\,x'^2\right]^{1-\lambda_0} 
                    \frac{1}{\tau^{\frac{3}{2}}}
                   \frac{\Delta^2}{8} (2 \tau)
                 \exp\!\left[\frac{3 \zeta^2 + 6\zeta \Delta + 4
                      \Delta^2}{2 \tau}\right] 
                  \nonumber\\
                   &&\!\!\!\!\!\!\!\!\!\times\!\!\left[2 (\zeta\!+\!2
                   \Delta)\exp\!\left[\!\frac{(\zeta\!+\!\Delta)^2}{\tau}\right]\!\!+\!
                  (\!(\zeta\!+\!2\Delta)^2 \!-\!2\tau)
                  \exp\!\left[\!\frac{5\zeta^2\!+\!12\zeta
                      \Delta\!+\!8\Delta^2}{4\tau}\right]\!\!\sqrt{\frac{\pi}{\tau}}
                  \mbox{Erf}\!\left[\!\frac{\zeta\!+\!2\Delta}{\sqrt{4
                        \tau}}\!\right]\!\right]. 
\label{eq_rhs_cr_3} \nonumber \\
\eea
Inserting this expression in (\ref{eq_T_lab_lam_2}), one obtains for
the scattering amplitude 
\bea
&&T(x,x',Y) = \frac{\alpha^2 x^2}{\lambda_0^2 (1-\lambda_0)^2}
                  \left[Q_c^2(Y)\,x'^2\right]^{1-\lambda_0} 
                    \frac{\Delta^2}{\tau^{\frac{1}{2}}}
                  \exp\!\left[\frac{3 \zeta^2 + 6\zeta \Delta + 4
                      \Delta^2}{2 \tau}\right] 
                  \nonumber\\
                   &&\!\!\!\!\!\!\!\!\!\times\!\!\left[2 (\zeta\!+\!2
                   \Delta)\exp\!\left[\!\frac{(\zeta\!+\!\Delta)^2}{\tau}\right]\!\!+\!
                  (\!(\zeta\!+\!2\Delta)^2 \!-\!2\tau)
                  \exp\!\left[\!\frac{5\zeta^2\!+\!12\zeta
                      \Delta\!+\!8\Delta^2}{4\tau}\right]\!\!\sqrt{\frac{\pi}{\tau}}
                  \mbox{Erf}\!\left[\!\frac{\zeta\!+\!2\Delta}{\sqrt{4
                        \tau}}\!\right]\!\right].
\label{eq_T_cr_4} \nonumber\\
\eea
This result is not the same as the result for the scattering amplitude
in laboratory frame (\ref{eq_T_fQ}). Thus, the Kovchegov equation or,
equivalently, the BFKL evolution in the presence of a saturation
boundary does not satisfy the completeness relation.

\subsubsection{Global saturation boundary}
\label{Sec_Global_saturation_boundary}
\begin{figure}[h!]
\setlength{\unitlength}{1.cm}
\begin{center}
\epsfig{file=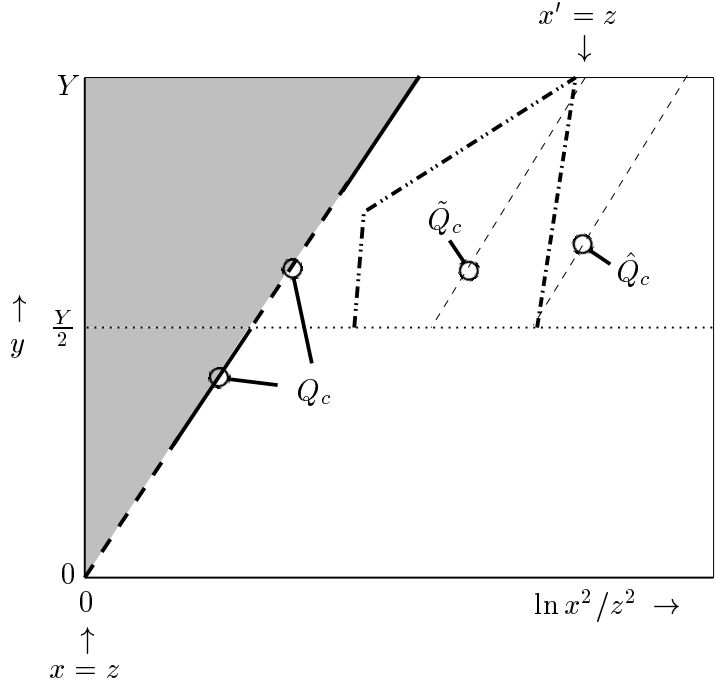, width=9.5cm}
\end{center}
\caption{BFKL evolution in two successive steps $0 \to Y/2 \to Y$ in
  the presence of a global saturation boundary $Q_c$.}
\label{Fig_BFKL_CR_GB}
\end{figure}

In this section we show that if the evolution of dipole $r_2$ from
$Y/2$ to $Y$ is limited by the saturation boundary $Q_c$ as shown in
Fig.~\ref{Fig_BFKL_CR_GB}, rather than $\hat{Q}_c$ as in the previous
section, then the scattering amplitude resulting from the Kovchegov
equation obeys the completeness relation.  However, in this case the
evolution of dipole $r_2$, for some values of $r_2$, violates
unitarity, i.e. $T(r_2,x',Y/2) > 1$.  To clarify this point, let's
consider the two regions:

$r_2 < 1/\tilde{Q}_c(Y/2)$: The amplitude
$T(r_2,x',Y/2)$ is always larger than $1$ since any evolution path
from $r_2$ to $x'$ goes through the internal saturation region of
dipole $r_2$ being on the lhs of $\hat{Q}_c$ as shown by the
dashed-dotted line in Fig.~\ref{Fig_BFKL_CR_GB}.  

$r_2 > 1/\tilde{Q}_c(Y/2)$: Each path of evolution from $x$ to $r_2$
to $x'$ respects unitarity.  However, also in this region there are
paths going from an intermediate point to another along which
$T(r_2,x',Y/2)$ becomes larger than $1$ due to the same reason as
above. For example along the dashed-dotted-dotted line in
Fig.~\ref{Fig_BFKL_CR_GB}.

The successive evolution $0 \to Y/2$ and $Y/2 \to Y$ in the presence
of a global (fixed) saturation boundary $Q_c$ as shown in
Fig.~\ref{Fig_BFKL_CR_GB} is given by
\bea
\mbox{[rhs of (\ref{eq_cr})]} &=& \frac{64}{\pi}\int \frac{d^2r_2}{2\pi r_2^2} 
\label{eq_rhs_cr_mean} \\ 
                 &&\hspace{-3cm}\times \left(\frac{x^2}{r_2^2}\right) 
                 \left[Q_c^2(Y/2)\,r_2^2\right]^{1-\lambda_0}
             \Delta \left[\ln\frac{1}{Q^2_c(Y/2)\,r_2^2} + \Delta\right]
             \exp\!\left[-\frac{\pi
                \ln^2(1/Q^2_c(Y/2)r_2^2)}{4 \alpha N_c
                \chi''(\lambda_0)Y/2} \right]
                \nonumber\\
                && \hspace{-3cm}\times \left( \frac{r_2^2}{x'^2}\right) 
             \left[\hat{Q}_c^2(Y/2)\,x'^2\right]^{1-\lambda_0}
             \left[\ln\!\left(\!\frac{Q_c^2(Y)r_2^2}{Q_c^2(Y/2)\,x^2}e^{-\Delta}\!\right)\, 
              \ln\!\left(Q_c^2(Y)\,x'^2 e^{-\Delta}\right) \right]
                \nonumber\\
                && \hspace{-3cm}\times
              \exp\!\left[-\frac{\pi
                \ln^2(1/\hat{Q}^2_c(Y/2)x'^2)}{4 \alpha N_c
                \chi''(\lambda_0)Y/2} \right] .\nonumber
\eea
The last two lines (multiplied by $8/\sqrt{\pi}$) give the dipole
number density $n(r_2,x',Y/2)$ which vanishes at $x'= e^{-\Delta}/Q_c$
or near the saturation boundary $Q_c$. To obtain this result, one
evaluates $n(r_2,x',Y)$ in the vicinity of $\hat{Q}_c$ following
\cite{Mueller:2002zm} and requires it to vanish at $x'=
e^{-\Delta}/Q_c(Y)$ (instead of $x'= e^{-\Delta}/\hat{Q}_c(Y)$) when
saturation (eq.(4.1) of \cite{Mueller:2002zm}) is taken into account.
This requirement opens ways of evolution which violate unitarity as
mentioned above. Note the difference between (\ref{eq_rhs_cr_mean}) and
(\ref{eq_rhs_cr}); in (\ref{eq_rhs_cr}) $n(r_2,x,Y/2)$ vanishes at
$x'= e^{-\Delta}/\hat{Q}_c$.

Using eq.~(\ref{eq_matching}) one gets 
\be
\ln\!\left(\!\frac{Q_c^2(Y)r_2^2}{Q_c^2(Y/2)\,x^2}e^{-\Delta}\!\right)
=    \ln(Q_c^2(Y/2)\,r_2^2) + \frac{3}{2(1-\lambda_c)}
\ln\left[\frac{\alpha N_c}{\pi}\chi''(\lambda_c) Y\right] - \Delta
\label{eq_rewr}
\ee
and
\be
\ln(1/\hat{Q}^2_c(Y/2)x'^2) = 
          \ln(Q_c^2(Y/2)\,r_2^2) + \frac{3}{2(1-\lambda_c)}
          \ln\left[\frac{\alpha N_c}{\pi}\chi''(\lambda_c) Y\right]
          - \ln(Q_c^2(Y)\,x'^2)\ 
\label{eq_rewr_1}
\ee
which allow us to evaluate eq.~(\ref{eq_rhs_cr_mean}) in the vicinity
of the saturation boundary $x'\approx 1/Q_c(Y)$ in the logarithmic
approximation. Using in addition also eq.(\ref{eq_def_z_e_t}), we get
\bea
\mbox{[rhs of (\ref{eq_cr})]} &=& - \frac{8}{\pi}\left(\frac{x^2}{x'^2}\right)
        \left[Q^2_c(Y) x'^2\right]^{1-\lambda_0}
        \frac{1}{\left[\frac{\alpha N_c}{\pi} \chi''(\lambda_0) Y\right]^{\frac{3}{2}}}
        \ln\!\left(Q_c^2(Y)\,x'^2 e^{-\Delta}\right)
\nonumber\\
        &&\hspace{-2cm}\times 
        \int_0^{\infty} d\ln(Q_c^2(Y/2)\,r_2^2)\
        \ln^3(Q_c^2(Y/2)\,r_2^2)\               
        \exp\!\left[-\frac{\pi
        \ln^2(Q^2_c(Y/2)r_2^2)}{\alpha N_c
        \chi''(\lambda_0)Y} \right] \ 
\eea
which after the integration over $\ln(Q_c^2(Y/2)\,r_2^2)$ gives 
\bea
\mbox{[rhs of (\ref{eq_cr})]} &=& \frac{8}{\pi}\left(\frac{x^2}{x'^2}\right)
        \left[Q^2_c(Y) x'^2\right]^{1-\lambda_0} \Delta 
        \left[\ln\frac{1}{Q_c^2(Y)\,x'^2} +\Delta\right] \ .
\eea
This result agrees with the result for the left-hand side of the
completeness relation given by (\ref{eq_n_ls}) (the additional
exponential function in (\ref{eq_n_ls}) becomes one in the very
vicinity of the saturation boundary which is considered here.). Note,
however, that in order to satisfy completeness one had to violate
unitarity.

\subsection{Frame-dependence}
\begin{figure}[h!]
\setlength{\unitlength}{1.cm}
\begin{center}
\epsfig{file=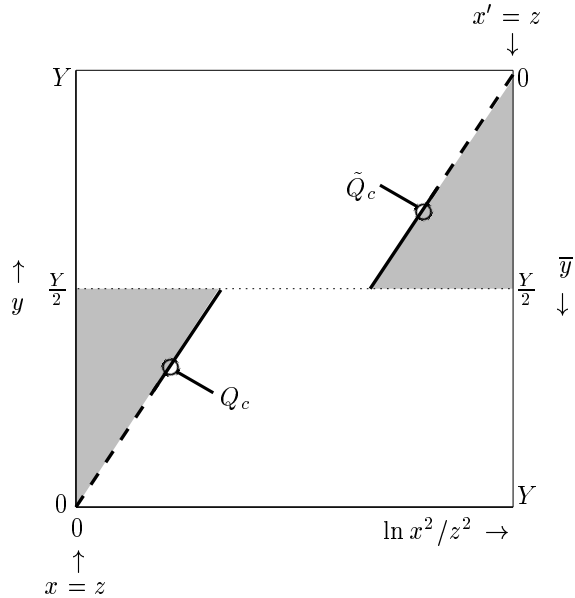, width=7.5cm}
\end{center}
\caption{Dipole-dipole scattering in the presence of saturation
  (shaded area) in the center-of-mass frame.}
\label{Fig_BFKL_CM}
\end{figure}

In this section we show that the result for the scattering amplitude
obtained in the laboratory frame (\ref{eq_T_fQ}) is frame-dependent.
For simplicity, we consider here the scattering of two dipoles in the
c.m.  frame (see Fig.~\ref{Fig_BFKL_CM}) and show that the
result disagrees with the result for the scattering amplitude in the
laboratory frame (\ref{eq_T_fQ}).

The scattering amplitude of two dipoles in the c.m. frame is given by
(\ref{eq_T_cm_1}), or after the integration over $r_1$, by
(\ref{eq_T_lab_lam}). With the dipole number densities
(\ref{eq_n_sat}) and (\ref{eq_n_sl_sat}), the scattering amplitude
(\ref{eq_T_lab_lam}) becomes 
\bea
T(x,x',Y) &=& \frac{\pi}{2} \alpha^2 \frac{1}{\lambda_0^2
    (1-\lambda_0)^2} \int\!\!\frac{d^2r_2}{2 \pi} \nonumber \\
    && \hspace{-3cm}\times \frac{8}{\sqrt{\pi}} \left(\frac{x^2}{r_2^2}\right) 
                 \left[Q_c^2(Y/2)\,r_2^2\right]^{1-\lambda_0}
             \Delta \left[\ln\frac{1}{Q^2_c(Y/2)\,r_2^2} + \Delta\right]
             \exp\!\left[\!-\frac{\pi
                \ln^2(1/Q^2_c(Y/2)r_2^2)}{4 \alpha N_c
                \chi''(\lambda_0)Y/2} \right]
                \nonumber\\
                && \hspace{-3cm}\times \frac{8}{\sqrt{\pi}}
            \left[\frac{1}{\tilde{Q}_c^2(Y/2)\,r_2^2}\right]^{1-\lambda_0}
            \Delta\left[\ln\tilde{Q}^2_c(Y/2)\,r_2^2 +\Delta\right]
            \exp\!\left[-\frac{\pi \ln^2(\tilde{Q}^2_c(Y/2)\,r_2^2)}{4 \alpha N_c
    \chi''(\lambda_0)Y/2}\right] \ , 
\label{eq_T_c_m_n} 
\eea
where $\tilde{Q}^2_c(Y/2)$ is given by (\ref{eq_Q_Qt}). For the dipole
of size $x$, the evolution proceeds from $y=0$ up to $y=Y/2$, while
for the dipole of size $x'$ evolution goes from $\overline{y}\equiv
Y-y=0$ to $\overline{y}=Y/2$ as shown in Fig.~\ref{Fig_BFKL_CM} where
the shaded saturation regions of the two dipoles are shown. With the
definitions in (\ref{eq_def_z_e_t}), one can easily see that the above
amplitude leads to the result (\ref{eq_T_cr_4}) which differs from the
result for the scattering amplitude in the laboratory frame
(\ref{eq_T_fQ}).

\section{BFKL evolution in the presence of saturation}
\ \label{BFKL_DW_SOL}

In this section we solve the BFKL equation in the presence of two
saturation boundaries, $\rho_1$ and $\rho_2$, as shown in
Fig.~\ref{Fig_BFKL_DW}. Consider the scattering process where dipole
$x$ evolves from $y=0$ up to $y=Y$ and then scatters on the elementary
dipole $x'$.  In this frame, the boundary $\rho_1$ is due to parton
saturation in the wavefunction of the evolved dipole $x$. The second
boundary $\rho_2$ excludes all paths of evolution from any
intermediate point ($\rho, 0<y<Y$) to the final point ($x'$, $Y$) with
$\rho>\rho_2$ which manifestly violate unitarity as discussed for the
case $r_2<1/\tilde{Q}_c(Y/2)$ in
sec.~\ref{Sec_Global_saturation_boundary}. In other words, the second
boundary $\rho_2$ ensures unitarity respecting amplitudes for each
path of evolution from ($x$, $0$) to ($x'$, $Y$) when the evolution
proceeds in two steps, from the initial point ($x$, $0$) to an
intermediate point ($\rho, y$) with $\rho_1<\rho<\rho_2$ and then from
the intermediate point ($\rho, y$) to the final point ($x'$, $Y$).
However, the second boundary does not eliminate unitarity violating
amplitudes when the evolution proceeds through two or more
intermediate stages; paths of evolution from an intermediate point to
another may violate unitarity since they may enter the internal
saturation regimes as explained in the case of
$r_2>1/\tilde{Q}_c(Y/2)$ in sec.~\ref{Sec_Global_saturation_boundary}.
To eliminate also this remaining unitarity violating evolution one has
to solve the Balitsky \cite{Balitsky:1995ub+X} or JIMWLK
\cite{Iancu:2001ad+X,Weigert:2000gi} equation which are difficult to
deal with analytically. The introduction of the second boundary
$\rho_2$ has another advantage: it makes the scattering amplitude
boost-invariant which was not the case for a single saturation
boundary. Thus, the calculation in the center of mass, say, now gives
the same result as the calculation in the laboratory frame of one of
the two dipoles.

We may view the scattering process also backwards, i.e., in a frame
where the evolved dipole $x'$ scatters on the elementary dipole $x$.
In this frame, the boundary $\rho_2$ is due to parton saturation in
the wavefunction of the evolved dipole $x'$ and $\rho_1$ excludes all
paths of evolution going to $x$ that manifestly violate unitarity.

\subsection{Diffusion in the presence of two absorptive barriers}
\label{sec_diffusion}
In this section we solve the diffusion equation in the presence of two
absorptive barriers and use the solution to calculate the scattering
amplitude in the following section.

Consider the diffusion equation
\be
\frac{\partial}{\partial t} \psi(\rho,t)=\frac{1}{4}
\frac{\partial^2}{\partial \rho^2} \psi(\rho,t) 
\label{eq_de_bc}
\ee
with the boundary conditions
\be
\psi(\rho_1,t) = \psi(\rho_2,t) = 0 
\label{eq_bc}
\ee
and the initial condition
\be
\psi(\rho,0) = \delta(\rho-\tilde{\rho}) \ .
\label{eq_ic}
\ee
This boundary-value problem can be solved by the method of separation
of variables. Specifically, we will (a) find solutions $\psi_n(\rho,t)
= f_n(\rho)g_n(t)$ of (\ref{eq_de_bc}) and (\ref{eq_bc}) and (b) find
the general solution $\psi(\rho,t)$ satisfying also the initial
condition (\ref{eq_ic}) by taking a suitable linear combination of the
functions $\psi_n(\rho,t)$.

(a) Inserting $\psi_n(\rho,t)= f_n(\rho)g_n(t)$ in (\ref{eq_de_bc}),
we obtain
\be
f_n(\rho)g'_n(t) = \frac{1}{4} f''_n(\rho) g_n(t)
\ee
or
\be
\frac{g'_n(t)}{g_n(t)/4} = \frac{f''_n(\rho)}{f_n(\rho)} \ .
\label{eq_t_rho}
\ee
Next, we observe that the left-hand side of (\ref{eq_t_rho}) is a
function of $t$ alone, while the right-hand side is a function of
$\rho$ alone. This implies that
\be
\frac{g'_n(t)}{g_n(t)/4} = \frac{f''_n(\rho)}{f_n(\rho)} = -\lambda_n
\label{eq_t_rho_lam}
\ee
for some constant $\lambda_n$. In addition, the boundary conditions
(\ref{eq_bc}) imply that $f_n(\rho_1)=0$ and $f_n(\rho_2)=0$. Hence
$\psi_n(\rho,t)$ is a solution of (\ref{eq_de_bc}) if
\be
f''_n(\rho) + \lambda_n f_n(\rho) = 0 \ ,\quad  f_n(\rho_1)=f_n(\rho_2)=0
\label{eq_bvp_f}
\ee
and
\be
g''_n(t) + \frac{1}{4} \lambda_n g_n(t)=0 \ .
\label{eq_bvp_g}
\ee
The boundary-value problem (\ref{eq_bvp_f}) has a solution only if
$\lambda_n = n^2 \pi^2/(\rho_2-\rho_1)^2$, and in this case, 
\be
f_n(\rho) = \sin\left(\frac{n\pi (\rho-\rho_1)}{\rho_2-\rho_1}\right) 
\ee
for every positive integer $n=1, 2 ...$ and $\rho_1 < \rho < \rho_2$.
Equation (\ref{eq_bvp_g}), in turn, implies that
\be
g_n(t) = e^{-\frac{n^2 \pi^2}{(\rho_2-\rho_1)^2}\frac{t}{4}} \ .
\ee
Hence,
\be
\psi_n(\rho,t) = c_n \sin\left(\frac{n\pi
    (\rho-\rho_1)}{\rho_2-\rho_1}\right)
   e^{-\frac{n^2 \pi^2}{(\rho_2-\rho_1)^2}\frac{t}{4}}
\ee
for every constant $c_n$ to be fixed by the initial condition (\ref{eq_ic}).
 
(b) The linear combination
\be
\psi(\rho,t) = \sum_{n=1}^{\infty}c_n \sin\left(\frac{n\pi
    (\rho-\rho_1)}{\rho_2-\rho_1}\right)
   e^{-\frac{n^2 \pi^2}{(\rho_2-\rho_1)^2}\frac{t}{4}}
\ee
satifies the boundary conditions (\ref{eq_de_bc}). To satisfy the initial
condition (\ref{eq_ic}), we must choose the constants $c_n$ such that
\be
\psi(\rho,0) = \sum_{n=1}^{\infty}c_n \sin\left(\frac{n\pi
    (\rho-\rho_1)}{\rho_2-\rho_1}\right)=\delta(\rho-\tilde{\rho}) \ .
\ee
on the interval $\rho_1 < \rho, \tilde{\rho} < \rho_2$. Using Fourier
analysis, the constants $c_n$ are
\be
c_n = \frac{2}{\rho_2-\rho_1} \int_{\rho_1}^{\rho_2}d\rho \
\delta(\rho-\tilde{\rho}) \sin\left(\frac{n\pi
    (\rho-\rho_1)}{\rho_2-\rho_1}\right) = 
\frac{2}{\rho_2-\rho_1} \sin\left(\frac{n\pi
    (\tilde{\rho}-\rho_1)}{\rho_2-\rho_1}\right) \ .
\ee
Thus, 
\be
\psi(\rho,t) = \sum_{n=1}^{\infty}
        \frac{2}{\rho_2-\rho_1} \sin\left(\frac{n\pi
        (\tilde{\rho}-\rho_1)}{\rho_2-\rho_1}\right)
        \sin\left(\frac{n\pi
        (\rho-\rho_1)}{\rho_2-\rho_1}\right)
        e^{-\frac{n^2 \pi^2}{(\rho_2-\rho_1)^2}\frac{t}{4}}
\label{eq_psi_f}
\ee
is the general solution.

\subsection{BFKL evolution between two saturation boundaries}
\label{Sec_BFKL_dw}

\begin{figure}[h!]
\setlength{\unitlength}{1.cm}
\begin{center}
\epsfig{file=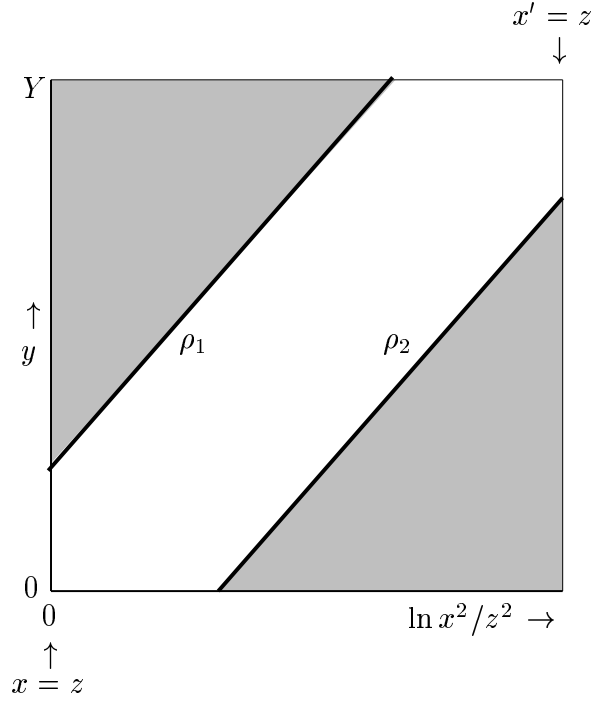, width=7.5cm}
\end{center}
\caption{BFKL evolution between two shaded saturation regimes.}
\label{Fig_BFKL_DW}
\end{figure}

Here we wish to evaluate the scattering amplitude in the laboratory frame
(\ref{eq_T_lab_lam}) in the region between the weak and the saturation
regimes in analogy to Ref.~\cite{Mueller:2002zm}.  Hence, we define a
particular line $Q_d^2(Y)$ in the $\ln(x^2/x'^2) - Y$ plane by the two
conditions
\be
\hspace{-2.8cm}\frac{2 \alpha N_c}{\pi}\chi'(\lambda_d)Y + \ln(Q_d^2\,x^2) = 0
\label{eq_sadpoi}
\ee
and
\be
\frac{2 \alpha N_c}{\pi}\chi(\lambda_d)Y -(1-\lambda_d)\ln(Q_d^2\,x^2) =
\frac{\pi^2}{(\rho_2-\rho_1)^2} \frac{\alpha
  N_c}{\pi}\chi''(\lambda_d) Y \ .
\label{eq_sadlin} 
\ee
Eq.~(\ref{eq_sadpoi}) is the saddle-point condition while
eq.~(\ref{eq_sadlin}) is chosen so that the scattering amplitude
becomes a constant along the line $Q_d(Y)$, as we will see. $\rho_2$
and $\rho_1$ denote the two boundaries limiting the BFKL evolution as
shown in Fig.~\ref{Fig_BFKL_DW}. The solution to (\ref{eq_sadpoi}) and
(\ref{eq_sadlin}) is
\be
(1-\lambda_d) \chi'(\lambda_d) + \chi(\lambda_d) = \frac{\pi^2}{2
  (\rho_2-\rho_1)^2}\chi''(\lambda_d)
\label{eq_lambda_d}
\ee
and
\be
Q^2_d(Y)\,x^2 = \exp\!\left[\frac{2\alpha
      N_c}{\pi}\frac{\chi(\lambda_d)}{1-\lambda_d}Y \left(1-\frac{\pi^2}{2
  (\rho_2-\rho_1)^2}\frac{\chi''(\lambda_d)}{\chi(\lambda_d)}\right)\right] .
\label{eq_Q_d}
\ee
Eq.~(\ref{eq_lambda_d}) determines $\lambda_d$ which is a constant for
fixed $Y$, $x$ and $x'$ or fixed $\rho_2-\rho_1$. For
large $\rho_2-\rho_1$, $\lambda_d$ approaches $\lambda_0=0.372$ which
is given by (\ref{eq_lambda_0}) since 
\be
\lambda_d-\lambda_0 = \frac{\pi^2}{2(\rho_2-\rho_1)^2} \frac{1}{1-\lambda_0}\ . 
\label{eq_lambda_ld_l0}
\ee
Now we evaluate the scattering amplitude~(\ref{eq_T_lab_lam}) in a
region around $Q^2_d(Y)$ by the saddle-point method, when $\alpha Y$
is large. This means that we expand the exponent in (\ref{eq_T_lab_lam}) around
$\lambda_d$ (since $\lambda_d$ satisfies the saddle-point condition)
and then use (\ref{eq_sadpoi}) and (\ref{eq_sadlin}). One obtains
\bea
\!\!\!\!\!\!\!T(x,x',Y)\!&\!=\!&\!\frac{\pi \alpha^2 x^2}{\lambda_d^2 (1-\lambda_d)^2} 
            \left(Q_d^2(Y)\,x'^2\right)^{1-\lambda_d} 
            \exp\!\left[\frac{\pi^2}{(\rho_2-\rho_1)^2} \frac{\alpha
            N_c}{\pi}\chi''(\lambda_d) Y \right]
            \nonumber \\
            &&\!\!\!\!\!\!\times \int \frac{d\lambda}{2\pi i} \exp\!\left[\frac{\alpha
                N_c}{\pi}\chi''(\lambda_d)Y(\lambda-\lambda_d)^2 + 
            (\lambda-\lambda_d)\ln\frac{1}{Q^2_d(Y)\,x'^2}\right] \ .
\eea
After the simple Gaussian integral, the amplitude becomes
\bea
T(x,x',Y) &=& \frac{\pi \alpha^2 x^2}{\lambda_d^2 (1-\lambda_d)^2} 
            \left(Q_d^2(Y)\,x'^2\right)^{1-\lambda_d} 
           \exp\!\left[\frac{\pi^2}{(\rho_2-\rho_1)^2} \frac{\alpha
            N_c}{\pi}\chi''(\lambda_d) Y \right]
            \nonumber\\
            &&\times \frac{1}{\sqrt{4 \alpha N_c
                \chi''(\lambda_d)Y}}\,\exp\!\left[-\frac{\pi
                \ln^2(1/Q^2_d(Y)\,x'^2)}{4 \alpha N_c
                \chi''(\lambda_d)Y}\right] \ .
\label{eq_T_all_Y}
\eea
To simplify notations, we use 
\be
\rho = \ln\left(\frac{x^2}{x'^2}\right), \quad
\rho_d(Y)=\ln(Q^2_d(Y)\,x^2), \quad t=\frac{4 \alpha N_c
  \chi''(\lambda_d)Y}{\pi}
\ee
and
\be
\psi(\rho-\rho_d,t) = \frac{1}{\sqrt{\pi t}}
                      e^{-\frac{(\rho-\rho_d)^2}{t}} \ .
\label{eq_wf}
\ee
In terms of the above the amplitude becomes
\be
T(x,x',Y) = \frac{\pi \alpha^2 x^2}{\lambda_d^2 (1-\lambda_d)^2} 
            e^{-(1-\lambda_d)(\rho-\rho_d)} \ 
           \exp\!\left[\frac{\pi^2}{(\rho_2-\rho_1)^2}
             \frac{t}{4}\right]\ 
            \psi(\rho-\rho_d,t) \ .
\label{eq_T_rt}
\ee
Here the function $\psi(\rho-\rho_d,t)$ represents the diffusive part of
the amplitude since it satisfies the diffusion equation
\be
\frac{\partial}{\partial t} \psi(\rho-\rho_d,t)=\frac{1}{4}
\frac{\partial^2}{\partial \rho^2} \psi(\rho-\rho_d,t) \ .
\label{eq_de}
\ee
Eq.(\ref{eq_wf}) is not the proper solution of the diffusion
equation~(\ref{eq_de}) in the presence of saturation. It is at this
point that we impose the unitarity. To obtain the right solution in
the presence of saturation, one has to exclude paths which go into the
saturation region while solving~(\ref{eq_de}).  This can be realized
by requiring $\psi(\rho-\rho_d,t)$ to vanish at the saturation
boundaries.  A solution to the diffusion equation with two saturation
boundary conditions is presented in Sec.~\ref{sec_diffusion}.
The general solution for $\psi(\rho-\rho_d,t)$ in the presence of
saturation is given by~(\ref{eq_psi_f}). However, to simplify
calculations, we will consider the scattering ampitude at large
rapidities,
\be
t \gg \frac{4(\rho_2-\rho_1)^2}{\pi^2} \ ,
\label{eq_t_large}
\ee
where the function $\psi(\rho-\rho_d,t)$ in (\ref{eq_psi_f}) is
dominated by the lowest mode $n=1$,
\be
\psi_s(\rho,t) = \frac{2}{\rho_2-\rho_1} \sin\left(\frac{\pi
        (\tilde{\rho}-\rho_1(0))}{\rho_2-\rho_1}\right)
        \sin\left(\frac{\pi
        (\rho-\rho_1)}{\rho_2-\rho_1}\right)
        e^{-\frac{\pi^2}{(\rho_2-\rho_1)^2}\frac{t}{4}} \ .
\label{eq_psi_lm}
\ee
Note that $\rho_1 < \rho < \rho_2$ (cf. Fig.~\ref{Fig_BFKL_DW}),
$\psi_s(\rho_1,t)=\psi_s(\rho_2,t)=0$, and $\tilde{\rho}$ is a
parameter which fixes the initial condition (\ref{eq_ic}); we choose
$\tilde{\rho}=0$ at $t=0$.

With (\ref{eq_psi_lm}) the scattering amplitude~(\ref{eq_T_rt}) now becomes
\be
T(x, x',Y)\!=\!\frac{\pi \alpha^2 x^2}{\lambda_d^2 (1\!-\!\lambda_d)^2} 
            \frac{2}{\rho_2\!-\!\rho_1} 
            e^{-(1-\lambda_d)(\rho-\rho_d(Y))}  
           \sin\!\left(\!\!\frac{\pi
           (\tilde{\rho}(0)\!-\!\rho_1(0))}{\rho_2-\rho_1}\!\!\right)
           \sin\!\left(\!\!\frac{\pi
           (\rho\!-\!\rho_1(Y))}{\rho_2-\rho_1}\!\!\right)\!.
\label{eq_T_dw}
\ee
In order to fix $\rho_2$ and $\rho_1$, we consider the amplitudes
$T(x,z,y)$ and $T(x',z,y)$ with $x>z>x'$ and $0<y<Y$. $T(x,z,y)$
follows from (\ref{eq_T_dw}) with $\rho$ replaced by
$\rho'=\ln(x^2/z^2)$ and $Y$ by $y$,
\be
T(x, z, y)\!=\!\frac{\pi \alpha^2 x^2}{\lambda_d^2 (1\!-\!\lambda_d)^2} 
            \frac{2}{\rho_2\!-\!\rho_1} 
            e^{-(1-\lambda_d)(\rho'-\rho_d(y))}  
           \sin\!\left(\!\frac{\pi
           (\tilde{\rho}(0)\!-\!\rho_1(0))}{\rho_2-\rho_1}\!\right)
           \sin\!\left(\!\frac{\pi
           (\rho'\!-\!\rho_1(y))}{\rho_2-\rho_1}\!\right)\!.
\label{eq_T_dw_x_z}
\ee
$T(x',z, y)$ can also be easily obtained; one starts with
(\ref{eq_T_lab_lam}), uses $\ln(1/\tilde{Q}^2_d(Y)\,x'^2)$ instead of
$\ln(Q_d^2(Y)\,x^2)$ in eqs.~(\ref{eq_sadpoi}) and (\ref{eq_sadlin}),
and follows otherwise the steps needed to compute the scattering
amplitude in (\ref{eq_T_dw}).  Following this way, one gets
\bea
\!\!\!\!\!\!\!\!T(x',z,y) &=&\frac{\pi \alpha^2 x^2}{\lambda_d^2 (1\!-\!\lambda_d)^2} 
            \frac{2}{\rho_2\!-\!\rho_1} 
            e^{(1-\lambda_d)(\rho'-\tilde{\rho}_d(y))}  
            \nonumber \\
            &&\!\!\!\times \sin\!\left(\!\frac{\pi
           (\rho_2(Y)-\ln(x^2/x'^2))}{\rho_2-\rho_1}\!\right)
           \sin\!\left(\!\frac{\pi
           (\rho_2(y)-\rho')}{\rho_2-\rho_1}\!\right)
\label{eq_T_dw_sl}
\eea
with 
\be
\tilde{\rho}_d(y)
%\!&=&\!\ln\left(\frac{x^2}{x'^2}\right) -
%                        \ln\left(\frac{1}{\tilde{Q}^2_d(y)x'^2}\right)
%\nonumber \\
                    = \ln\left(\frac{x^2}{x'^2}\right) -
                        \frac{2\alpha
      N_c}{\pi}\frac{\chi(\lambda_d)}{1-\lambda_d} \left(1-\frac{\pi^2}{2
  (\rho_2-\rho_1)^2}\frac{\chi''(\lambda_d)}{\chi(\lambda_d)}\right)(Y-y) .
\label{eq_tilde_rho}
\ee
The amplitude $T(x,z,y)$ becomes maximum,
\be
\left. \frac{\partial}{\partial \rho'}T(x,z,y)\right|_{\rho_{is}} = 0 \ ,
\ee
at the saturation boundary,
\be
\rho_{is}(y) = \rho_1(y) + \frac{1}{1-\lambda_d} \ .
\label{eq_Q_s_dw}
\ee
We emphasize that here $\rho_{is}$ is the saturation line for internal
evolution once $x$, $x'$ and $Y$ have been fixed. It is not the
saturation line for our overall scattering matrix $T(x, x', Y)$. We
determine $\rho_1(y)$ by setting the amplitude at $\rho_{is}$
\be
T(\rho' = \rho_{is},y) = 2 \pi x^2 c,
\label{eq_T_unit_0}
\ee
with $c$ a constant of order $1$ as required by unitarity.
Eqs.~(\ref{eq_Q_s_dw}) and (\ref{eq_T_unit_0}) give
\be
\rho_1(y)\!=\!\frac{2\alpha
             N_c}{\pi}\frac{\chi(\lambda_d)}{1\!-\!\lambda_d} 
            \!\left[1\!-\!\frac{\pi^2}{2
          (\rho_2\!-\!\rho_1)^2}\frac{\chi''(\lambda_d)}{\chi(\lambda_d)}\right]\!y  
             - \frac{1}{1\!-\!\lambda_d} \ln\!\left[\!\frac{c
               \lambda_d^2 (1\!-\!\lambda_d)^3 (\rho_2\!-\!\rho_1)^2
               e}{\pi \alpha^2 \sin\left(-\pi \rho_1(0)/(\rho_2\!-\!\rho_1)\right)}\right]\!.
\label{eq_rho1}
\ee
Using $T(x',z,y)$ given by (\ref{eq_T_dw_sl}) instead
of $T(x,z,y)$, one analogously obtains
\be
\tilde{\rho}_{is}(y) = \rho_2(y) - \frac{1}{1-\lambda_d} 
\label{eq_tQ_s_dw}
\ee
and
\bea
\!\!\!\!\!\!\!\!\!\rho_2(y) =  \ln\left(\frac{x^2}{x'^2}\right) 
           \!\! &+&\!\! \frac{1}{1-\lambda_d} \ln\left[\frac{c
               \lambda_d^2 (1-\lambda_d)^3 (\rho_2-\rho_1)^2
               e}{\pi \alpha^2 \sin\left(\frac{\pi(\rho_2(Y)-\ln(x^2/x'^2))}{\rho_2-\rho_1}\right)}\right]
            \nonumber \\
            \!\!&-&\!\!\frac{2\alpha
             N_c}{\pi}\frac{\chi(\lambda_d)}{1-\lambda_d} 
            \left[1-\frac{\pi^2}{2
          (\rho_2-\rho_1)^2}\frac{\chi''(\lambda_d)}{\chi(\lambda_d)}\right](Y-y)\ . 
\label{eq_rho2}
\eea
Thus,
\bea
\!\!\!\!\!\rho_2-\rho_1 = \ln\left(\frac{x^2}{x'^2}\right)
             \!&-&\! \frac{2\alpha
             N_c}{\pi}\frac{\chi(\lambda_d)}{1-\lambda_d} 
            \left[1-\frac{\pi^2}{2
            (\rho_2-\rho_1)^2}\frac{\chi''(\lambda_d)}{\chi(\lambda_d)}\right]\,Y
        \nonumber\\
            &+&  \frac{2}{1-\lambda_d} \ln\left[\frac{c
                \lambda_d^2 (1-\lambda_d)^3 (\rho_2-\rho_1)^2
               e}{\pi \alpha^2 \sin\left(-\pi \rho_1(0)/(\rho_2-\rho_1)\right)}\right]
\label{eq_rho2_rho1_com}
\eea
or in terms of the saturation line
\be
\rho_2-\rho_1 = \ln\left(\frac{x^2}{x'^2}\right) - \rho_{is}(Y) + 
                \frac{1}{1-\lambda_d} \ln\left[\frac{c
                \lambda_d^2 (1-\lambda_d)^3 (\rho_2-\rho_1)^2
               e^2}{\pi \alpha^2 \sin\left(-\pi \rho_1(0)/(\rho_2-\rho_1)\right)}\right]
\label{eq_rho2_rho1}
\ee
with 
\bea
\rho_{is}(Y) &=& \frac{2\alpha
             N_c}{\pi}\frac{\chi(\lambda_d)}{1-\lambda_d} 
            \left[1-\frac{\pi^2}{2
            (\rho_2-\rho_1)^2}\frac{\chi''(\lambda_d)}{\chi(\lambda_d)}\right]\,Y  
           + \frac{2}{1-\lambda_d}
           \nonumber\\
          &&\hspace{2cm} - \frac{1}{1-\lambda_d} \ln\left[\frac{c
             \lambda_d^2 (1-\lambda_d)^3 (\rho_2-\rho_1)^2
             e^2}{\pi \alpha^2 \sin\left(-\pi
             \rho_1(0)/(\rho_2-\rho_1)\right)}\right] \ .
\label{eq_rho_s}
\eea
The above quantities $\rho_1$, $\rho_2$, $\rho_2-\rho_1$, and $\rho_{is}$ are
determined once the dipole sizes $x$ and $x'$ and their relative rapidity $Y$ are known.

The distance $\rho_2-\rho_1$ becomes minimum in the region
$\ln(x^2/x'^2) \ge \rho_{is}(Y)$ exactly at $\ln(x^2/x'^2) =
\rho_{is}(Y)$ or $x' = 1/Q_{is}(Y)$, in which case (\ref{eq_rho2_rho1})
reduces to
\be
\Delta\rho = \frac{1}{1-\hat{\lambda}_d} \ln\left[\frac{c
               \hat{\lambda}_d^2 (1-\hat{\lambda}_d)^3 (\Delta\rho)^2
               e^2}{\pi \alpha^2 \sin\left(-\pi \hat{\rho}_1(0)/\Delta\rho\right)}\right]
\label{eq_deltarho}
\ee
where $\Delta\rho$ denotes the minimum of $\rho_2-\rho_1$ while
$\hat{\lambda}_d$ and $\hat{\rho}_1(0)$ follow from (\ref{eq_lambda_d}) and
 (\ref{eq_rho1}) with $\rho_2-\rho_1$ replaced by $\Delta\rho$.
Comparing (\ref{eq_rho1}) with (\ref{eq_deltarho}), we have
\be
-\hat{\rho}_1(0) = \Delta\rho -\frac{1}{1-\hat{\lambda}_d}
\ee
which when inserted in (\ref{eq_deltarho}) leads to
\be
\Delta\rho = \frac{1}{1-\hat{\lambda}_d} \ln\left[\frac{c
               \hat{\lambda}_d^2 (1-\hat{\lambda}_d)^4 (\Delta\rho)^3
               e^2}{\pi^2 \alpha^2 }\right] \ .
\label{eq_deltarho_1}
\ee
This equation can be solved by iteration, for small $\alpha$, and the
solution is
\be
\Delta\rho(\alpha) = \frac{2}{1-\lambda_0}\ln\frac{1}{\alpha} +
                     \frac{3}{1-\lambda_0}\ln\ln\frac{1}{\alpha} +
                     \mbox{$\cal{O}$(const.)} \ .
\ee

The scaling behavior of $T$, as given in (\ref{eq_T_dw}), and even the
value of the saturation momentum is rather subtle. At first glance it
appears that (\ref{eq_T_dw}) would give an exponential behavior in
$\rho$, a behavior identical to the scaling behavior known to follow
from the Kovchegov equation. However, this is not the case because
$\rho_d(Y)$ has a strong $\rho$-dependence also. To investigate
further the $\rho$-dependence of $T$ define
\be
\frac{d\rho_d(Y)}{d\rho} \equiv \zeta
\label{eq_der_rho_d}
\ee
where the $\rho$ and $Y$ dependence of $\zeta$ is to be
determined. Differentiating (\ref{eq_sadpoi}) with respect to $\rho$
gives
\be
\frac{d\lambda_d}{d\rho} = \frac{-\zeta}{2 \alpha N_c
    \chi''(\lambda_d) Y/\pi} \approx \frac{-\zeta}{2 \alpha N_c
    \chi''(\lambda_0) Y/\pi}
\label{eq_der_lam}
\ee
while differentiating (\ref{eq_lambda_ld_l0}) gives
\be
\frac{d\lambda_d}{d\rho} = - \frac{\pi^2}{(1-\lambda_0)(\rho_2 -
  \rho_1)^3} \frac{d(\rho_2-\rho_1)}{d\rho} \ .
\label{eq_der_lam_1}
\ee
Equating the right hand side of (\ref{eq_der_lam}) and
(\ref{eq_der_lam_1}) leads to
\be
\frac{d(\rho_2-\rho_1)}{d\rho} = \frac{\zeta (1-\lambda_0)
  (\rho_2-\rho_1)^3}{2 \pi \alpha N_c \chi''(\lambda_0) Y} \ .
\label{eq_der_rho21}
\ee
We may also use (\ref{eq_rho1}) and (\ref{eq_rho2}) to evaluate the
$\rho$-dependence of $\rho_1$ and $\rho_2$. It is straightforward to
see that
\be
\frac{d\rho_1(Y)}{d\rho} = \zeta + {\cal
  O}\left(\frac{1}{\rho_2-\rho_1}\frac{d(\rho_2-\rho_1)}{d\rho}
\right)
\label{eq_der_rho1}
\ee
and
\be
\frac{d\rho_2(Y)}{d\rho} = 1 + {\cal
  O}\left(\frac{1}{\rho_2-\rho_1}\frac{d(\rho_2-\rho_1)}{d\rho}
\right) \ .
\label{eq_der_rho2}
\ee
Thus
\be
\frac{d(\rho_2(Y)-\rho_1(Y))}{d\rho} \equiv 
\frac{d(\rho_2-\rho_1)}{d\rho} = 1 - \zeta + {\cal
  O}\left(\frac{1}{\rho_2-\rho_1}\frac{d(\rho_2-\rho_1)}{d\rho}
\right) \ .
\label{eq_der_rho21_1}
\ee

Dropping the small term of the right hand side of
(\ref{eq_der_rho21_1}) and using (\ref{eq_der_rho21}), one finds
\be
\frac{d(\rho_2-\rho_1)}{d\rho} =
\frac{(1-\lambda_0)(\rho_2-\rho_1)^3}{2 \pi \alpha N_c
  \chi''(\lambda_0) Y} \left[1+ \frac{(1-\lambda_0)(\rho_2-\rho_1)^3}{2 \pi \alpha N_c
  \chi''(\lambda_0) Y} \right]^{-1} \ .
\label{eq_der_rho21_sf}
\ee
While an exact solution of (\ref{eq_der_rho21_sf}) looks difficult, it
is easy to see the general features of the solution.

When $\rho$ is not too large the solution to (\ref{eq_der_rho21_sf}),
neglecting the final factor $[\,\,]^{-1}$, is
\be
\rho_2-\rho_1 = \frac{\Delta\rho}{\sqrt{1-\frac{(1-\lambda_0)(\rho_2-\rho_1)^2}{\pi \alpha N_c
  \chi''(\lambda_0) Y} \left(\rho-\hat{\rho}_1(Y)-\frac{1}{1-\hat{\lambda}_d}\right)}} \ .
\label{eq_rho21_res}
\ee
Eq.~(\ref{eq_rho21_res}) is valid until $\rho_2-\rho_1$ becomes on the
order of $(\alpha Y)^{1/3}$ at which point the character of the
$\rho$-dependence changes and $\zeta \to 0$ giving
\bea
&&\rho_1(Y) \approx \mbox{const}' 
\label{eq_rho1_res} \\ 
&&\rho_2(Y) \approx \rho + \mbox{const}'' \ .
\label{eq_rho2_res}
\eea
Also, since
\be
\frac{d(\rho-\rho_d(Y))}{d\rho} = 1-\zeta \approx
\frac{d(\rho_2-\rho_1)}{d\rho}
\label{eq_rho_rho_d}
\ee
one has
\be
\rho-\rho_d(Y) = \rho_2-\rho_1+\mbox{const} 
\label{eq_rho_rho_d_res}
\ee
so that $T$, as given by (\ref{eq_T_dw}), stays very close to the
value $2 \pi x^2 c$ starting at $\rho =
\hat{\rho}_1(Y)+\frac{1}{1-\hat{\lambda}_d}$ up to $\rho=
\hat{\rho}_1(Y)+\frac{1}{1-\hat{\lambda}_d}+{\cal O}\left(\frac{\pi \alpha N_c
    \chi''(\lambda_0)Y}{(1-\lambda_0)(\Delta \rho)^3}\right)$. This
means that the saturation momentum for the scattering of the dipole
$x'$ on the dipole $x$ at rapidity $Y$ is 
\bea
\rho_s(Y) &=&   \hat{\rho}_1(Y) + \frac{1}{1-\hat{\lambda}_d} + 
     c_s \frac{\pi \alpha N_c
    \chi''(\lambda_0)Y}{(1-\lambda_0)(\Delta \rho)^3} \nonumber\\
          &=& \hat{\rho}_{is}(Y) + c_s \frac{\pi \alpha N_c
    \chi''(\lambda_0)Y}{(1-\lambda_0)(\Delta \rho)^3}
\label{eq_rho_s_reg}
\eea
with $c_s$ a matter of choice. While (\ref{eq_rho_s_reg}) indicates,
parametrically, a small percentage deviation from the $\rho_s$ that
occurs using the Kovchegov equation that deviation is very
significant. The $c_s$ in (\ref{eq_rho_s_reg}) is determined by the
exact value of $T$ at which one declares saturation to have occured.
Varying the value of $T$ by, say, $2$ will vary $c_s$ by a similar
factor which also means that $T$ is a function of
$[\rho-\rho_s(Y)]/(\alpha Y/(\Delta\rho)^3)$ and does not have the
usual scaling behavior of the Kovchegov equation.

Finally, it is a little curious, and perhaps unexpected, that one now
has a very extended region in $\rho$ where $T$ is near its unitarity
limit and where our formalism applies. This is not the case with the
Kovchegov equation. As $\rho$ varies from $\hat{\rho}_1(Y)$ to
$\rho_s(Y)$ the interval $\rho_2-\rho_1$ hardly changes and so the
description of $T$ hardly changes. What is happening is that the
evolution from $y=0$ to $y=Y$, for a given $\rho$, has a fairly well
defined scale of gluons at each intermediate value of $y$ which
contribute to the evolution. When $\rho$ changes the scale of the
intermediate gluons also changes, but the form of the dynamics remains
exactly the same. In this picture unitarity limits involve a rather
well defined scale of gluons at each value of $y$, but that scale
depends on the initial and final dipole sizes.

\subsection{Consistency checks}
In this section we show that out result for the scattering amplitude
(\ref{eq_T_dw}) is boost-invariant. If the saturation
boundaries which are fixed by $x$, $x'$ and $Y$ are used for any
evolution step in rapidity between $0$ and $Y$ (global saturation
boundaries as in sec.~\ref{Sec_Global_saturation_boundary}), then the
amplitude (\ref{eq_T_dw}) obeys also the completeness relation.

Let us start with the completeness relation (\ref{eq_cr}). The dipole
number density on the left-hand side of (\ref{eq_cr}) can be
calculated analogous to the scattering amplitude in the previous
section and in the case where $x>x'$ it is
\bea
\!\!\!\!\!n(x,x',Y) &=& \frac{4}{\rho_2-\rho_1} 
             \left(\frac{x^2}{x'^2}\!\right) 
             \left[Q_d^2(Y)\,x'^2\right]^{1-\lambda_d}
             \nonumber \\
           &&\times
           \sin\left(\frac{\pi
           (\tilde{\rho}(0)-\rho_1(0))}{\rho_2-\rho_1}\right)
           \sin\left(\frac{\pi
           (\ln(x^2/x'^2)-\rho_1(Y))}{\rho_2-\rho_1}\right) \ .
\label{eq_n_dw_ls}
\eea
Using this expression, the right-hand side of (\ref{eq_cr}) becomes
\bea
&&\!\!\!\!\!\!\!\mbox{[rhs of (\ref{eq_cr})]} =
                        \frac{16}{(\rho_2-\rho_1)^2} 
                         \int \frac{d^2r_2}{2\pi r_2^2} 
\label{eq_rhs_cr_dw_2} \\ 
            &&\!\!\!\!\!\!\!\times 
             \left(\frac{x^2}{r_2^2}\right) 
             \left[Q_d^2(Y/2)\,r_2^2\right]^{1-\lambda_d}
           \sin\left(\frac{\pi
           (\tilde{\rho}(0)-\rho_1(0))}{\rho_2-\rho_1}\right)
           \sin\left(\frac{\pi
           (\ln(x^2/r_2^2)-\rho_1(Y/2))}{\rho_2-\rho_1}\right)
           \nonumber \\
            &&\!\!\!\!\!\!\!\times 
            \left(\frac{r_2^2}{x'^2}\right)\!\! 
            \left[\hat{Q}_d^2(Y/2)\,x'^2\right]^{1-\lambda_d}
           \sin\!\left(\!\frac{\pi
           (\ln(x^2/r_2^2)\!-\!\rho_1(Y/2))}{\rho_2-\rho_1}\right)
           \sin\!\left(\!\frac{\pi
           (\ln(x^2/x'^2)\!-\!\rho_1(Y))}{\rho_2-\rho_1}\!\right),
           \nonumber
\label{eq_rhs_dw_sb}
\eea
where $Q^2_d(Y/2)$ is given by (\ref{eq_Q_d}), with $Y$ replaced by
$Y/2$, and $\hat{Q}^2_d(Y/2)$ is determined by matching the the two evolution
steps at $Y/2$,
\be
\hat{Q}_d^2(Y/2)\,r_2^2 = Q_d^2(Y/2)\,x^2 \ .
\ee
This equation and $\xi = \ln(x^2/r_2^2)$ allow us to write the
amplitude (\ref{eq_rhs_cr_dw_2}) as 
\bea
\!\!\mbox{[rhs of (\ref{eq_cr})]} &=&
                        \frac{16}{(\rho_2-\rho_1)^2} 
            \left(\frac{x^2}{x'^2}\right)
            \left[Q_d^2(Y)\,x'^2\right]^{1-\lambda_d}
           \nonumber \\
            &&\!\!\!\!\!\!\!\times 
           \sin\left(\frac{\pi
           (\tilde{\rho}(0)-\rho_1(0))}{\rho_2-\rho_1}\right)
           \sin\!\left(\!\frac{\pi
           (\ln(x^2/x'^2)\!-\!\rho_1(Y))}{\rho_2-\rho_1}\!\right)
           \nonumber \\
            &&\!\!\!\!\!\!\!\times 
           \int_{\rho_1(Y/2)}^{\rho_2(Y/2)} \frac{d\xi}{2} 
           \sin^2\left(\frac{\pi
           (\xi-\rho_1(Y/2))}{\rho_2-\rho_1}\right)
            \nonumber
\label{eq_cr_dw_3}
\eea
which after the integration over $\xi$,
\be
           \int_{\rho_1(Y/2)}^{\rho_2(Y/2)} \frac{d\xi}{2} 
           \sin^2\left(\frac{\pi
           (\xi-\rho_1(Y/2))}{\rho_2-\rho_1}\right)
           = 
           \frac{\rho_2-\rho_1}{4},
\label{eq_xi_int}
\ee
agrees with (\ref{eq_n_dw_ls}). Thus, our formalism satisfies the
completeness relation.

Now let us calculate the scattering amplitude of two dipoles in the
c.m. frame starting with (\ref{eq_T_cm_1}). To do this, in addition to
$n(x,r_2,Y/2)$ given by (\ref{eq_n_dw_ls}), we need also
the dipole number density $n(x', r_2, Y/2)$ with $x'<r_2$. The later is
obtained in the same way as the scattering amplitude (\ref{eq_T_dw_sl}) in
the previous section and it is 
\bea
n(x',r_2,Y/2) &=& \frac{4}{\rho_2-\rho_1}  
                \left[\frac{1}{\tilde{Q}_d^2(Y/2)\,r_2^2}\right]^{1-\lambda_d}
\label{eq_n_dw_sl}  \\
           &&\times \sin\left(\frac{\pi
           (\rho_2(Y)-\ln(x^2/x'^2))}{\rho_2-\rho_1}\right)
           \sin\left(\frac{\pi
           (\rho_2(Y/2)-\ln(x^2/r^2_2))}{\rho_2-\rho_1}\right) \nonumber
\eea
where $\tilde{Q}^2_d(Y)$ is determined by
\be
 Q^2_d(Y)\,x'^2 = \frac{1}{\tilde{Q}^2_d(Y)\,x^2} 
\label{eq_Q_rel_d}
\ee
with $Q^2_d(Y)$ given in (\ref{eq_Q_d}). Inserting the dipole number
densities in (\ref{eq_T_lab_lam}), the scattering amplitude in c.m. frame becomes 
\bea
&&T(x, x', Y) =  \frac{\pi \alpha^2}{2\lambda_d^2 (1-\lambda_d)^2} \frac{16}{(\rho_2-\rho_1)^2} 
                 \int \frac{d^2r_2}{2\pi} 
\label{eq_T_cm_dw} \\ 
            &&\!\!\!\!\!\!\!\times 
             \left(\frac{x^2}{r_2^2}\right) 
             \left[Q_d^2(Y/2)\,r_2^2\right]^{1-\lambda_d}
           \sin\left(\frac{\pi
           (\tilde{\rho}(0)-\rho_1(0))}{\rho_2-\rho_1}\right)
           \sin\left(\frac{\pi
           (\ln(x^2/r_2^2)-\rho_1(Y/2))}{\rho_2-\rho_1}\right)
           \nonumber \\
            &&\!\!\!\!\!\!\!\times 
            \left[\frac{1}{\tilde{Q}_d^2(Y/2)\,r_2^2}\right]^{1-\lambda_d}
            \sin\left(\frac{\pi
           (\rho_2(Y)-\ln(x^2/x'^2))}{\rho_2-\rho_1}\right)
           \sin\left(\frac{\pi
           (\rho_2(Y/2)-\ln(x^2/r^2_2))}{\rho_2-\rho_1}\right) \ .
           \nonumber
\eea
With (\ref{eq_Q_rel_d}), the variable $\xi = \ln(x^2/r_2^2)$ and the
equality 
\bea
            \sin\left(\frac{\pi
           (\rho_2(Y)-\ln(x^2/x'^2))}{\rho_2-\rho_1}\right)\!\!\!
           &&\!\!\!\sin\left(\frac{\pi
           (\rho_2(Y/2)-\ln(x^2/r^2_2))}{\rho_2-\rho_1}\right) =
           \\
            &&\hspace{-2cm}\sin\left(\frac{\pi
           (\ln(x^2/x'^2)-\rho_1(Y))}{\rho_2-\rho_1}\right)
           \sin\left(\frac{\pi
           (\ln(x^2/r^2_2)-\rho_1(Y/2))}{\rho_2-\rho_1}\right) \ , \nonumber
 \eea
the amplitude (\ref{eq_T_cm_dw}) reads
\bea
T(x, x', Y) &=&  \frac{\pi \alpha^2 x^2}{2\lambda_d^2 (1-\lambda_d)^2} 
                 \frac{16}{(\rho_2-\rho_1)^2} 
             \left[Q_d^2(Y)\,x'^2\right]^{1-\lambda_d}
\label{eq_T_cm_dw_xi} \\ 
            &&\times 
           \sin\left(\frac{\pi
           (\tilde{\rho}(0)-\rho_1(0))}{\rho_2-\rho_1}\right)
            \sin\left(\frac{\pi
           (\ln(x^2/x'^2)-\rho_1(Y))}{\rho_2-\rho_1}\right)
            \nonumber \\
            &&\times 
           \int_{\rho_1(Y/2)}^{\rho_2(Y/2)} \frac{d\xi}{2} 
           \sin^2\left(\frac{\pi
           (\xi - \rho_1(Y/2))}{\rho_2-\rho_1}\right) \ .
            \nonumber
\eea
After the integration over $\xi$ (see eq.~(\ref{eq_xi_int})), the
resulting scattering amplitude agrees with the result for the
scattering amplitude obtained in the laboratory frame (\ref{eq_T_dw}).

% ___ Running Coupling _________________
\section{BFKL evolution with running coupling and \\ boundaries}
\label{Sec_Running_Coupling}
% ____________________________________________________________________________
We now add a running coupling to the discussion of
Sec.~\ref{BFKL_DW_SOL}. It is difficult to use a global version of
BFKL evolution, covering all $Y$-values, such as given in
(\ref{eq_T_all_Y}), so we shall instead use a more local approach. The
boundaries $\rho_1$ and $\rho_2$ of Fig.~\ref{Fig_BFKL_DW} are now no
longer straight lines, however, the problem can still be formulated in
terms of a similar picture with curved boundaries to be
determined. BFKL evolution goes from $x$ at $y=0$ to $x'$ at
$y=Y$. Between $\rho_1$ and $\rho_2$ the normal BFKL kernel is used,
but we demand a solution which vanishes on the boundaries. The
scattering amplitude thus depends on $x$, $x'$ and $Y$, the variables
fixing the region of evolution. In order to do the actual calculation
additional variables $z$ and $y$ are introduced. $z$ and $y$ are the
dipole size and rapidity of an intermediate stage in the evolution
between $x$ and $x'$. The BFKL evolution is written in terms of $z$
and $y$. In fixed coupling evolution these intermediate stages of
evolution were not so visible because we were able to identify the
part of the BFKL equation that corresponded to diffusion, and then it
was possible to combine the results of global BFKL evolution, as given
in (\ref{eq_T_rt}), with the solution to the diffusion equation with
boundaries, given in (\ref{eq_psi_f}). We note, however, that $T$ as
given in (\ref{eq_T_dw}) does not obey BFKL evolution, say in the
variables $x'$ and $Y$, because a variation of these variables changes
the boundaries. The solution (\ref{eq_T_dw}) represents evolution
between $x$ at $y=0$ and $x'$ at $y=Y$ for fixed $x$, $x'$ and $Y$ and
it is this problem that we now turn to in case the coupling runs.

With the variables $x$, $x'$ and $Y$ being fixed we consider the
scattering amplitude $T(z, y)$ going from $x$ at $y=0$ to $z$ at $y$,
obeying the BFKL equation, and vanishing on the boundaries
$\rho_1(y)$, $\rho_2(y)$. Because running of the coupling introduces
an explicit scale it is convenient to now use variables scaled by
$\Lambda^2$ rather than by $x^2$. Thus in this section our notation is
\bea
\rho_i &=& \ln\left(1/\Lambda^2 x^2\right), \quad \rho_f=\ln\left(1/\Lambda^2 x'^2\right), \quad \rho=\ln\left(1/\Lambda^2 z^2\right)
\nonumber \\
\rho_d &=& \ln\left(Q_d^2/\Lambda^2\right), \quad \ \rho_1=\ln\left(Q_1^2/\Lambda^2\right), \quad \rho_2=\ln\left(Q_2^2/\Lambda^2\right)
\eea
where $Q_1$ and $Q_2$ are the momenta setting the boundaries, and
$\rho_i$ and $\rho_f$ represent the initial and final dipole sizes.

The BFKL equation takes the form
\be
\frac{\partial}{\partial y} \left[T/\alpha\right] = \frac{\alpha
  N_c}{\pi} K \times \left[T/\alpha\right]
\label{eq_BFKL_eq}
\ee
with $K$ the usual BFKL kernel. The convolution indicated in
(\ref{eq_BFKL_eq}) can be written more fully as 
\be
\left[K \times f\right](\rho) = \int_{-\infty}^\infty d\rho' K(\rho,
\rho') f(\rho') \ .
\label{eq_K_1}
\ee
The BFKL kernel is given as 
\be
K(\rho, \rho') = \int \frac{d\lambda}{2 \pi i}\,
e^{-(1-\lambda)(\rho-\rho')}\,2 \chi(\lambda)
\label{eq_K_2}
\ee
and we look for a solution to (\ref{eq_BFKL_eq}) where $T(\rho, y)$
vanishes when $\rho = \rho_1, \rho_2$. The running coupling is given,
as usual, by
\be
\alpha(\rho) = \frac{1}{b \rho} \ .
\label{eq_run_coup}
\ee
The problem of finding a solution to (\ref{eq_BFKL_eq}) with $K$ given
in (\ref{eq_K_2}) and where the solution vanishes at $\rho_1$ and
$\rho_2$ is a well defined mathematical problem, but it appears too
difficult to solve exactly. However, following
Refs.~\cite{Mueller:2002zm} and \cite{Camici:1996fr} we
can get an approximate solution to the problem by expanding
$\chi(\lambda)$ in (\ref{eq_K_2}) about $\lambda_d$
\be
\chi(\lambda) \approx \chi(\lambda_d)
+(\lambda-\lambda_d)\,\chi'(\lambda_d) + \frac{1}{2}
(\lambda-\lambda_d)^2\, \chi''(\lambda_d)
\ee
in which case
\be
K(\rho, \rho') \approx e^{-(1-\lambda_d) \rho} \, 2 \left[
  \chi(\lambda_d) + \chi'(\lambda_d)\frac{\partial}{\partial \rho} +
  \frac{1}{2} \chi''(\lambda_d)\,\frac{\partial^2}{\partial
    \rho^2}\right] e^{(1-\lambda_d)\,\rho} \delta(\rho-\rho')
\label{eq_K_lam}
\ee
with $\lambda_d$ to be determined much as was done in Sec.~\ref{Sec_BFKL_dw}.

Using (\ref{eq_K_lam}) in (\ref{eq_BFKL_eq}) gives
\be
\frac{\partial}{\partial y} \left[T/\alpha\right](\rho,y) = \frac{2 N_c}{\pi b
  \rho}\,e^{-(1-\lambda_d) \rho}\!\!\left[\!
  \chi(\lambda_d)\!+\!\chi'(\lambda_d)\frac{\partial}{\partial \rho} +
  \frac{1}{2} \chi''(\lambda_d)\,\frac{\partial^2}{\partial
    \rho^2}\!\right]\!e^{(1-\lambda_d)\,\rho}
\left[T/\alpha\right](\rho,y).
\label{eq_BFKL_app}
\ee
There are three distinct regions of parameters determining how, and to
what extent, (\ref{eq_BFKL_app}) can be solved:
\bea
&&(\mbox{i}) \quad \frac{\rho_2-\rho_1}{Y^{1/6}} \gg 1 
\label{eq_gg_1}\\
&&(\mbox{ii}) \quad \frac{\rho_2-\rho_1}{Y^{1/6}} \ll 1 
\label{eq_ll_1}\\
&&(\mbox{iii}) \quad \frac{\rho_2-\rho_1}{Y^{1/6}} \approx 1 \ .
\label{eq_app_1}
\eea
In region (i) expand the $1/\rho$ coming from the running coupling in
(\ref{eq_BFKL_app}) as 
\be
\frac{1}{\rho} = \frac{1}{\rho_d} - \frac{(\rho-\rho_d)}{\rho_d^2} 
\label{eq_rho_exp}
\ee
where, as in Sec.~\ref{Sec_BFKL_dw}, $\rho_d(y)=\rho$ defines a line
of constant $T$. Then (\ref{eq_rho_exp}) leads to diffusion in a
linear potential. Condition (i) corresponds to an exponentially small
amplitude near the boundary $\rho_2$ and the problem reduces to the
problem of solving (\ref{eq_BFKL_app}) with a single boundary at
$\rho_1$ which has the solution given in
Ref.~\cite{Mueller:2002zm}. In region (ii) one may replace $1/\rho$ by
$1/\rho_d$ since the range $\rho_1<\rho<\rho_2$ is so small that the
linear potential has very little effect. In this region we can
effectively use a fixed coupling equation locally in $y$, but one must
take into account that $\rho_d$ depends on $y$. We shall shortly carry
this out in detail. It appears difficult to get an analytic solution
in region (iii), but the form of the solution is strongly constraint
by the known solution in region (i) and (ii).

Now we turn to the solution  of (\ref{eq_BFKL_app}) in region (ii)
where $\rho$ can be replaced by $\rho_d$. Since we expect the
$\rho$-dependence of the solution to be the same, in form, to the
fixed coupling case it is natural to write
\be
T/\alpha = \sin\left(\frac{\pi (\rho-\rho_1)}{\rho_2-\rho_1}\right)
f(\rho, y) \ .
\label{eq_T_ans}
\ee
In addition we suppose $\rho_2-\rho_1$ is very weakly dependent on $y$
so that (\ref{eq_BFKL_app}) gives, with $\dot{\rho}_1 = d\rho_1/dy$,
\bea
\hspace{-0.7cm}\!\!\!&&-\frac{\pi \dot\rho_1}{\rho_2-\rho_1} \cos\left(\frac{\pi
    (\rho-\rho_1)}{\rho_2-\rho_1}\right) f + \sin\left(\frac{\pi
    (\rho-\rho_1)}{\rho_2-\rho_1}\right) \frac{\partial f}{\partial y}=
\label{eq_sin} \\
\hspace{-0.7cm}&& \frac{2 N_c}{\pi b
  \rho_d}\,e^{-(1-\lambda_d)(\rho-\rho_d)}\!\!\left[
  \chi(\lambda_d) + \chi'(\lambda_d)\frac{\partial}{\partial \rho} +
  \frac{1}{2} \chi''(\lambda_d)\,\frac{\partial^2}{\partial
    \rho^2}\right]\!e^{(1-\lambda_d) (\rho-\rho_d)} \sin\left(\frac{\pi
    (\rho-\rho_1)}{\rho_2-\rho_1}\right) f .\nonumber
\eea
Eq.~(\ref{eq_T_dw}) suggests that the $\rho$-dependence of $f$ should
be $e^{-(1-\lambda_d)(\rho-\rho_d)+\gamma \ln y}$, with the $\gamma
\ln y$ term due to the fact that (\ref{eq_sin}) is an equation for
$T/\alpha$ not for $T$, in which case (\ref{eq_sin}) gives 
\be
-\dot\rho_1 = \frac{2N_c}{\pi b \rho_d} \chi'(\lambda_d)
\label{eq_rho_der}
\ee
from the $\cos\left(\frac{\pi (\rho-\rho_1)}{\rho_2-\rho_1}\right)$
part of the equation, and 
\be
\frac{\gamma}{y} + (1-\lambda_d)\dot\rho_d = \frac{2 N_c}{\pi b \rho_d}
\left[\chi(\lambda_d) - \frac{1}{2} \chi''(\lambda_d)
  \frac{\pi^2}{(\rho_2-\rho_1)^2} \right]
\label{eq_sin_part}
\ee
coming from the
$\sin\left(\frac{\pi(\rho-\rho_1)}{\rho_2-\rho_1}\right)$ part of the
equation. The $\gamma/y$ term in (\ref{eq_sin_part}) is small and can
be dropped. Since $\rho_d$ defines a line of constant amplitude
$\dot\rho_d$ and $\dot\rho_1$ must be exactly equal in which case
(\ref{eq_rho_der}) and (\ref{eq_sin_part}) becomes identical to
(\ref{eq_sadpoi}) and (\ref{eq_sadlin}) when the form of the running
coupling, (\ref{eq_run_coup}), is taken into account.

Eqs.~(\ref{eq_rho_der}) and (\ref{eq_sin_part}) give (\ref{eq_lambda_d}) along
with 
\be
\frac{d}{dy} \rho_d^2 = \frac{4 N_c}{\pi b}
\frac{\chi(\lambda_d)}{1-\lambda_d} \left[ 1- \frac{1}{2}
  \frac{\pi^2}{(\rho_2-\rho_1)^2}
  \frac{\chi''(\lambda_d)}{\chi(\lambda_d)}\right]
\label{eq_rho_deriv}
\ee
or
\be
\rho_d(y) = \sqrt{\frac{4 N_c}{\pi b}
\frac{\chi(\lambda_d)}{1-\lambda_d} \left[ 1- \frac{1}{2}
  \frac{\pi^2}{(\rho_2-\rho_1)^2}
  \frac{\chi''(\lambda_d)}{\chi(\lambda_d)}\right] y} + c_d \ .
\label{eq_rho_d_rc}
\ee
The constant $c_d$ in (\ref{eq_rho_d_rc}) follows not from solving
(\ref{eq_rho_deriv}), but reflects the fact that we have not
systematically kept terms in (\ref{eq_sin}) and in (\ref{eq_rho_deriv})
 of size $1/y$ and $1/\sqrt{y}$, respectively. One such term is the
 $\gamma/y$ term in (\ref{eq_sin_part}). Taking $\alpha(\rho) \approx
 [b(\rho_d+c')]^{-1}$ rather than $\alpha(\rho) \approx
 [b\rho_d]^{-1}$ would also give a contribution to $c_d$. In contrast
 to Ref.~\cite{Mueller:2002zm}, here we do not have control over such
 terms. As before (\ref{eq_lambda_d}) gives $\lambda_d$, approximately
 as in (\ref{eq_lambda_ld_l0}), when $\rho_2-\rho_1$ is large. Since
 $\rho_d$ has the same $y$-dependence as $\rho_1$ and $\rho_2$ we can
 write
\be
\rho_{1,2}= \sqrt{\frac{4 N_c}{\pi b}
\frac{\chi(\lambda_d)}{1-\lambda_d} \left[ 1- \frac{1}{2}
  \frac{\pi^2}{(\rho_2-\rho_1)^2}
  \frac{\chi''(\lambda_d)}{\chi(\lambda_d)}\right] y} + c_{1,2} \ .
\label{eq_rho_12_rc}
\ee
where $c_d$ and $c_1$, $c_2$ can depend on $\rho_i$ and $\rho_f$. From
(\ref{eq_T_ans}) and taking $\gamma=1/2$ to guarantee the lack of
$y$-dependence of $T$ when $\rho-\rho_d$ is fixed, one can write
\be
T(\rho,y) = 2 \pi x^2 c \frac{(\rho_2-\rho_1)}{\pi} e (1-\lambda_d)
\sin\left(\frac{\pi(\rho-\rho_1)}{\rho_2-\rho_1}\right)
e^{-(1-\lambda_d)(\rho-\rho_1(y))} \ .
\label{eq_T_run_c}
\ee
$T$, as given in (\ref{eq_T_run_c}) and with $\rho_1$ given in
(\ref{eq_rho_12_rc}), satisfies the BFKL equation (\ref{eq_BFKL_eq})
as well as the boundary conditions. In addition
\be
T(\rho_1(y)+\frac{1}{1-\lambda_d}, y) = 2 \pi x^2 c \ .
\label{eq_T_unit}
\ee
Eq.~(\ref{eq_T_unit}) is a saturation, or unitarity, condition for
evolution at intermediate values of the rapidity.

Eq.~(\ref{eq_T_run_c}) is not so strong as our fixed coupling
equation (\ref{eq_T_dw}) which gives the amplitude for evolution from a given
dipole of size $x$ in terms of boundary lines $\rho_1$ and $\rho_2$
which then are completely determined in terms of the evolution. In the
present case we are able to do the evolution at large rapidities, but
we are unable to connect to the initial dipole, or
hadron. Nevertheless, there is only one extra unknown constant here as
compared to the fixed coupling case as we now demonstrate. First, we
note that the relationship between $\rho_2(Y)$ and $\rho_f$ is here
essentially as it was in the fixed coupling case. In the fixed
coupling case $\rho_2-\rho_f$ was determined by symmetry from the
determination of $\rho_1-\rho_i$ which followed from
unitarity. However, we equally well could have determined $\rho_2$ by
unitarity in which case the condition is
\be
T(\rho=\rho_2(y)-\frac{1}{1-\lambda_d}, \rho_f, y) = 2 \pi r_2^2 c \ .
\label{eq_T_unit_1}
\ee
where
\be
r_2^2 = \frac{1}{\Lambda^2} e^{-(\rho_2(y)-\frac{1}{1-\lambda_d})} \ .
\label{eq_r_2}
\ee
Eqs.~(\ref{eq_T_unit_1}) and (\ref{eq_r_2}) determine the curve
$\rho_2(y)$ as the right-most value of $\rho$ from which BFKL
evolution respects the unitarity limit in going from $\rho$ to
$\rho_f$. Thus at the maximum rapidity (\ref{eq_rho2}), except for
some notational changes required in the running coupling case,
remains valid. That is
\be
\rho_2(Y) = \rho_f + \frac{1}{1-\lambda_d} \ln\left[\frac{c
               \lambda_d^2 (1-\lambda_d)^3 (\rho_2-\rho_1)^3
               e}{\pi^2 \alpha^2 (\rho_2(Y)-\rho_f)}\right] \ .
\label{eq_rho2_rc}
\ee
Eq.~(\ref{eq_rho2_rc}) determines the constant $c_2$ in
(\ref{eq_rho_12_rc}) to be
\be
c_2 = \rho_f - \sqrt{\frac{4 N_c}{\pi b}
\frac{\chi(\lambda_d)}{1\!-\!\lambda_d}\!\!\left[1- \frac{1}{2}
  \frac{\pi^2}{(\rho_2-\rho_1)^2}
  \frac{\chi''(\lambda_d)}{\chi(\lambda_d)}\right]Y} + 
\frac{1}{1\!\!-\!\!\lambda_d} \ln\!\left[\frac{c
               \lambda_d^2 (1-\lambda_d)^3 (\rho_2\!-\!\rho_1)^3
               e}{\pi^2 \alpha^2 (\rho_2(Y)-\rho_f)}\right]\!.
\ee
However, we are not able to determine $c_1$. Determining $c_1$ requires
evolving into regions where (\ref{eq_ll_1}) is no longer valid and
hence would require an ability to solve the BFKL equation keeping the
running of the coupling as given in (\ref{eq_T_ans}). We do not know
how to do this.

One can get some insight into the regions demarcated by
(\ref{eq_gg_1})-(\ref{eq_app_1}) by looking at $\rho_d(y)$ as given by
(\ref{eq_rho_d_rc}) as one reduces $y$ to the point where
$\frac{\rho_2-\rho_1}{y^{1/6}}$ is no longer so small. We can write
\be
\rho_d(y) \approx \sqrt{\frac{4 N_c}{\pi b}
\frac{\chi(\lambda_d)}{1-\lambda_d} y } \left[ 1- \frac{1}{4}
  \frac{\pi^2}{(\rho_2-\rho_1)^2}
  \frac{\chi''(\lambda_d)}{\chi(\lambda_d)}\right] \ .
\ee
When $\frac{\rho_2-\rho_1}{y^{1/6}}$ is of order $1$, $\rho_d$ has a
dominant $\sqrt{y}$ term along with a term of size $y^{1/6}$. But,
this is exactly what happens in the region (\ref{eq_gg_1}) as seen in
the solution of Ref.~\cite{Mueller:2002zm}. Thus when $\frac{\rho_2-\rho_1}{y^{1/6}}$ is
of order $1$ there are $y^{1/6}$ corrections coming both from the
boundaries and from the linear potential. When
$\frac{\rho_2-\rho_1}{y^{1/6}} \gg 1$ the linear potential effects
dominate the effects due to the $\rho_2$ boundary. When
$\frac{\rho_2-\rho_1}{y^{1/6}} \ll 1$ the boundary effects dominate
the linear potential effects.

%----------------------------------------------

\section*{Acknowledgements}

A.~Sh. acknowledges financial support by the Deutsche Forschungsgemeinschaft under
contract Sh 92/1-1.

%
% ___ Bibliography __________________________________________________
%
  
%
%
\end{document}